\documentclass[prl,twocolumn,superscriptaddress]{revtex4-2}
\usepackage{amssymb}
\usepackage{amsfonts}
\usepackage{bbm}
\usepackage{mathrsfs}
\usepackage{graphics,graphicx,epsfig,bm,amsmath,amsthm,amssymb}
\usepackage{bm}
\usepackage{bbm}
\usepackage{longtable}
\usepackage{multirow}
\usepackage{array}
\usepackage{color}
\usepackage{cases}
\usepackage{booktabs }
\usepackage[usenames,dvipsnames]{xcolor}

\usepackage{xcolor}
\usepackage[none]{hyphenat}
\usepackage{float}
\usepackage{amsmath}
\usepackage[a4paper,colorlinks=true,
linkcolor=blue,citecolor=blue,
pdfauthor={ },
pdftitle={ },
pdfsubject={ },
pdfkeywords={ }]{hyperref}

\newcommand{\R}[1]{\uppercase\expandafter{\romannumeral#1}}

\begin{document}

\title{Scalable architecture for dark photon searches: Superconducting-qubit proof of principle}

\author{Runqi Kang}
\affiliation{CAS Key Laboratory of Microscale Magnetic Resonance and School of Physical Sciences, University of Science and Technology of China, Hefei 230026, China}
\affiliation{Anhui Province Key Laboratory of Scientific Instrument Development and Application, University of Science and Technology of China, Hefei 230026, China}
\affiliation{Hefei National Laboratory, University of Science and Technology of China, Hefei 230088, China}

\author{Qingqin Hu}
\affiliation{CAS Key Laboratory of Microscale Magnetic Resonance and School of Physical Sciences, University of Science and Technology of China, Hefei 230026, China}
\affiliation{Anhui Province Key Laboratory of Scientific Instrument Development and Application, University of Science and Technology of China, Hefei 230026, China}

\author{Xiao Cai}
\affiliation{Gusu Laboratory of Materials, Suzhou, China}

\author{Wenlong Yu}
\affiliation{Gusu Laboratory of Materials, Suzhou, China}

\author{Jingwei Zhou}
\email{zhoujw@ustc.edu.cn}
\affiliation{CAS Key Laboratory of Microscale Magnetic Resonance and School of Physical Sciences, University of Science and Technology of China, Hefei 230026, China}
\affiliation{Anhui Province Key Laboratory of Scientific Instrument Development and Application, University of Science and Technology of China, Hefei 230026, China}
\affiliation{Hefei National Laboratory, University of Science and Technology of China, Hefei 230088, China}

\author{Xing Rong}
\email{xrong@ustc.edu.cn}
\affiliation{CAS Key Laboratory of Microscale Magnetic Resonance and School of Physical Sciences, University of Science and Technology of China, Hefei 230026, China}
\affiliation{Anhui Province Key Laboratory of Scientific Instrument Development and Application, University of Science and Technology of China, Hefei 230026, China}
\affiliation{Hefei National Laboratory, University of Science and Technology of China, Hefei 230088, China}

\author{Jiangfeng Du}
\affiliation{CAS Key Laboratory of Microscale Magnetic Resonance and School of Physical Sciences, University of Science and Technology of China, Hefei 230026, China}
\affiliation{Anhui Province Key Laboratory of Scientific Instrument Development and Application, University of Science and Technology of China, Hefei 230026, China}
\affiliation{Hefei National Laboratory, University of Science and Technology of China, Hefei 230088, China}
\affiliation{Institute of Quantum Sensing and School of Physics, Zhejiang University, Hangzhou 310027, China}

\date{\today}

\begin{abstract}
	The dark photon is a well-motivated candidate of dark matter due to its potential to open the window of new physics beyond the Standard Model.
    A fundamental mass-range-sensitivity dilemma is always haunting the dark photon searching experiments: The resonant haloscopes have excellent sensitivity but are narrowband, and vice versa for the non-resonant ones.
    A scalable architecture integrating numerous resonant haloscopes will be a desirable solution to this dilemma. However, even the concept of scalable searching remains rarely explored, due to the size limitation of conventional haloscopes imposed by the dark photon wavelength.
	Here we propose and demonstrate a novel architecture using superconducting qubits as sub-wavelength haloscope units. By virtue of the scalability of superconducting qubits, it is possible to integrate multiple qubits with different frequencies on a chip-scale device. Furthermore, the frequencies of the qubits can be tuned to extend the searching mass range. Thus, our architectures allow for searching for dark photons in a broad mass range with high sensitivity.
	As a proof-of-principle experiment, we designed and fabricated a three-qubit chip and successfully demonstrated a scalable dark-photon searching.  Our work established  constraints on dark photons in the mass range of 15.632 µeV$\sim$15.638 µeV, 15.838 µeV$\sim$15.845 µeV, and 16.463 µeV$\sim$16.468 µeV, simultaneously, and the constraints are much more stringent than the cosmology constraints. Our work can be scaled up in the future to boost the scrutiny of new physics and extended to search for more dark matter candidates, including dark photons, axions and axion-like particles.
\end{abstract}

\maketitle

The hypothetical dark matter shines a light on the ultimate secret of what the universe is made up of \cite{1906Poincare,1987Trimble,2010Feng,2018Bertone}.
Dark matter also holds great potential to explain many experimental and astronomical anomalies \cite{1933Zwicky,2009Cholis,2021Cazzaniga,2022Thomas}.
The simplest $U(1)$ gauge boson dark photon is a prime candidate for dark matter. An especially compelling motivation of dark photons comes from the development of theories beyond the Standard Model, such as string theories and unified field theories.
Since many compelling theories assume large symmetry groups, the smallest $U(1)$ symmetry can naturally arise from the decomposition of these high-rank groups \cite{1986Holdom1,1986Holdom2,2009Goodsell,2009Abazov,2021Searight}.
Dark photons are predicted to couple very weakly to ordinary matter and have a mass in the range of $10^{-22}$ eV $-$ 1 eV \cite{2023Kimball}.
Such vast parameter spaces make the experimental search for dark photons a huge challenge.

Experimental searches for dark photons rely on their kinetic mixing with ordinary photons.
One of the most widely utilized technique is the haloscope \cite{2020Filippi,2021Caputo,2023Kimball}.
Projects, such as HAYSTAC \cite{2017Brubaker}, CAPP \cite{2021Kwon}, QUAX \cite{2021Alesini,2022Alesini}, ADMX \cite{2022Cervantes1}, Dark SRF \cite{2023Romanenko}, ORGAN \cite{2024Quiskamp}, and SHANHE \cite{2024Tang}, use microwave cavities operating at a few gigahertz to conduct resonant detections at µeV scale.
The bandwidths of the cavities are usually several hundred kilohertz, depending on the quality factor.
While experiments like FUNK \cite{2020Andrianavalomahefa}, BREAD \cite{2022Liu}, DOSUE \cite{2023Kotaka}, QUALIPHIDE \cite{2023Ramanathan}, and BRASS-p \cite{2023Bajjali} are non-resonant haloscopes, which have larger bandwidths (typically a few gigahertz) at the expense of lower sensitivity.
The SQuAD group employed a superconducting qubit to read out the cavity, and achieved a sub-standard-quantum-limit sensitivity \cite{2021Dixit}.
In a subsequent work of SQuAD, they prepared the cavity in a Fock state via a superconducting qubit, and realized a 2.78-fold enhancement on the signal rate \cite{2024Agrawal}.
For the purpose of suppressing thermal noise, most haloscope experiments are carried out in cryogenic systems, such as dilution refrigerators.
Given the vast potential parameter space of dark photons, a scalable architecture integrating plentiful resonant haloscopes covering a broad mass range can be beneficial to the search for dark photons.
However, despite intense efforts devoted to the search for dark photons, the scalability of haloscopes remains rarely investigated.
Present haloscopes rely on devices with a size comparable to the wavelength of dark photons (on the order of 0.1 m), while the limited space in cryogenic systems usually hinders the haloscope devices to scale up.

Superconducting qubits have emerged as the most promising structure for various applications in the booming field of quantum information due to their remarkable scalability \cite{2004Blais,2023Acharya}.
Most superconducting qubits operate at a few gigahertz, while their size is typically on the micrometer scale, much smaller than the corresponding wavelength.
The small size of superconducting qubits allows integrating a great number of superconducting qubits in a chip-scale device.
In this article, we describe a novel architecture based on superconducting qubits for experimental searching for dark photons.
The superconducting qubits are utilized as sensitive dark photon detectors.
Each qubit can be resonantly driven by a dark photon field with a frequency close to the eigenfrequency of the qubit.
By integrating numerous qubits with various eigenfrequencies, simultaneously searching for dark photons in a wide frequency range can be realized.
The searching mass range can be further extended by tuning the frequencies of the qubits through magnetic flux.
We fabricated a three-qubit chip to demonstrate this proposal.
Our experiment successfully shows that the search for dark photon can be conducted with three tunable frequencies simultaneously. Thus, the scalability of the proposal has been verified. Our result established constraints on the kinetic mixing of dark photons in the mass ranges of 15.632 µeV$\sim$15.638 µeV, 15.838 µeV$\sim$15.845 µeV, and 16.463 µeV$\sim$16.468 µeV, surpassing the constraints from CMB distortion by 1$\sim$2 orders of magnitude.
\begin{figure}[http]
	\centering
	\includegraphics[width=1\columnwidth]{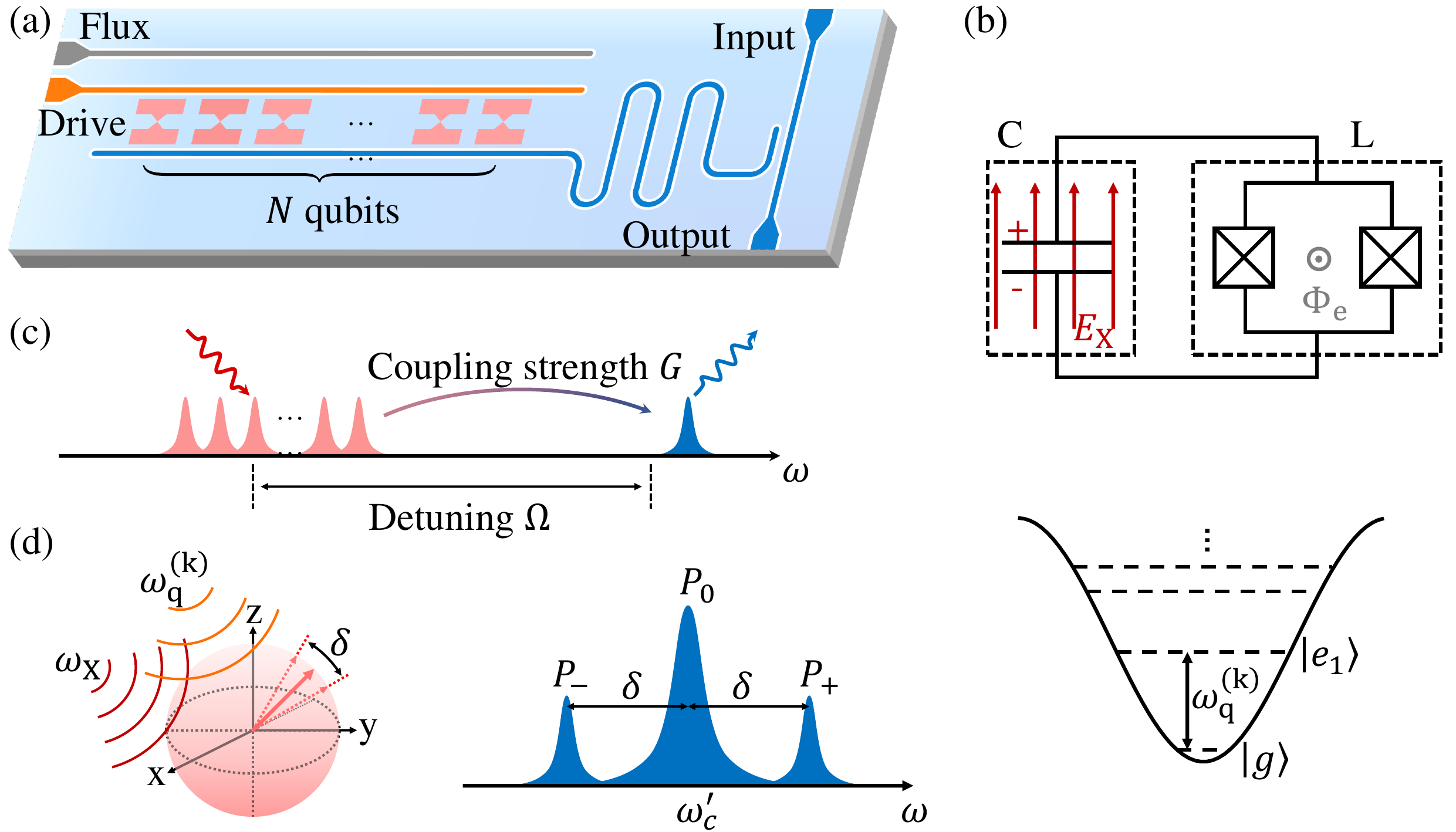}
	\caption{
		The dark photon searching scheme.
		(a) The scalable superconductor architecture for dark photon search.
			It consists of $N$ superconducting transmon qubits (light coral), a readout cavity (blue), a drive line (dark orange), and a flux line (gray).
		(b) Circuit diagram (up) and energy levels (down) of a transmon qubit.
			The qubit is equivalent to an LC oscillator.
			The capacitance is sensitive to external electric field (red), while the inductance is sensitive to external magnetic flux (gray).
			The non-linearity of the oscillator results in a non-quadratic potential (solid line) and unevenly spaced energy levels (dashed lines).
		(c) Illustration of the concept of many-to-one mapping. All qubits are dispersively coupled to the cavity, so that the output of the cavity contains information at all frequencies of the qubits.
		(d) Principle of heterodyne detection.
        The dark orange curves refer to the driving microwave with frequency $\omega_{\rm q}^{\rm (k)}$ and the dark red curves refer to the dark photon field with frequency $\omega_{\rm X}$. There are three peaks in the output of the cavity. The central peak with power $P_{0}$ is the reflected probe microwave, while the sideband peaks with the power $P_{\pm}$ and detuning $\delta = |\omega_{\rm X}-\omega_{\rm q}^{\rm (k)}|$ are the dark-photon-induced signal.}
	\label{Figure1}
\end{figure}

Fig. \ref{Figure1}(a) shows the architecture for scalable dark photon searching, which contains an array of $N$ superconducting transmon qubits, a resonator, a drive line, and a flux line.
Figure \ref{Figure1}(b) shows the circuit diagram and energy levels of a transmon qubit.
Each qubit is equivalent to an LC circuit with a capacitance $C$ made of two conductor plates and an inductance $L$ made of two Josephson junctions.
The nonlinear nature of Josephson junctions gives rise to a cosine potential energy term and leads to unevenly spaced energy levels. Consequently, even though a qubit has infinite energy levels, one can regard the qubit as a two-level system with the eigenfrequency $\omega_{\rm q}^{\rm (k)}$, where $\textrm{k} = 1,\dots N$ is the index of the qubit.
If the qubit is in a dark photon field, the capacitance would sense the dark-photon-induced electric field \cite{2023Chen},
\begin{equation}
    E_{\rm X} = \kappa\chi\sqrt{2\rho_{\rm DM}/\epsilon_{0}},
\end{equation}
where $\chi$ is the kinetic mixing, $\rho_{\rm DM}$ is the local dark matter density \cite{2014Read}, $\epsilon_{0}$ is the vacuum permittivity, and $\kappa\sim\mathcal{O}(1)$ is the package coefficient introduced by the metallic surroundings of the qubits.
The two Josephson junctions form a superconducting quantum interference device (SQUID). The inductance $L$ of the SQUID is dependent on the magnetic flux through it, $\Phi_{\rm e}$, so the eigenfrequency of the qubit can be tuned by changing the magnetic flux.
Figure \ref{Figure1}(c) depicts the concept of the many-to-one mapping from the detecting frequencies to the readout frequency.
The eigenfrequencies of the qubits are designed slightly different from each other, forming a comb-like structure.
The qubits couple to the readout resonator in the dispersive regime, i.e., the detuning $\Omega_{\rm c}^{\rm (k)} = |\omega_{\rm c}-\omega_{\rm q}^{\rm (k)}|$ is larger than the coupling $G^{\rm (k)}$. The Hamiltonian of the interaction between the k-th qubit and the cavity is
\begin{equation}
	\widehat{\mathcal{H}}_{\rm int}^{\rm (k)} = \hbar g^{\rm (k)}\widehat{\tau}_{\rm z}^{\rm (k)}\widehat{a}^{\dagger}\widehat{a},
\end{equation}
where $\widehat{\tau}_{\rm z}^{\rm (k)}$ is the Pauli operator of the k-th qubit, $g^{\rm (k)} = G^{\rm(k)2}/\Omega_{\rm c}^{\rm (k)}$, and $\widehat{a}^{\dagger}$, $\widehat{a}$ are the creation and annihilation operators of the cavity \cite{2004Blais}.
For an arbitrary qubit, the interaction will transduce the change of $\widehat{\tau}_{\rm z}^{\rm (k)}$ into a shift of the resonant frequency of the cavity.
Since each qubit detects the signal resonantly, and the qubit array covers a wide frequency range, searching for dark photons in a broad frequency range simultaneously with high sensitivity is feasible through our architectures.

\begin{figure}[http]
	\centering
	\includegraphics[width=1\columnwidth]{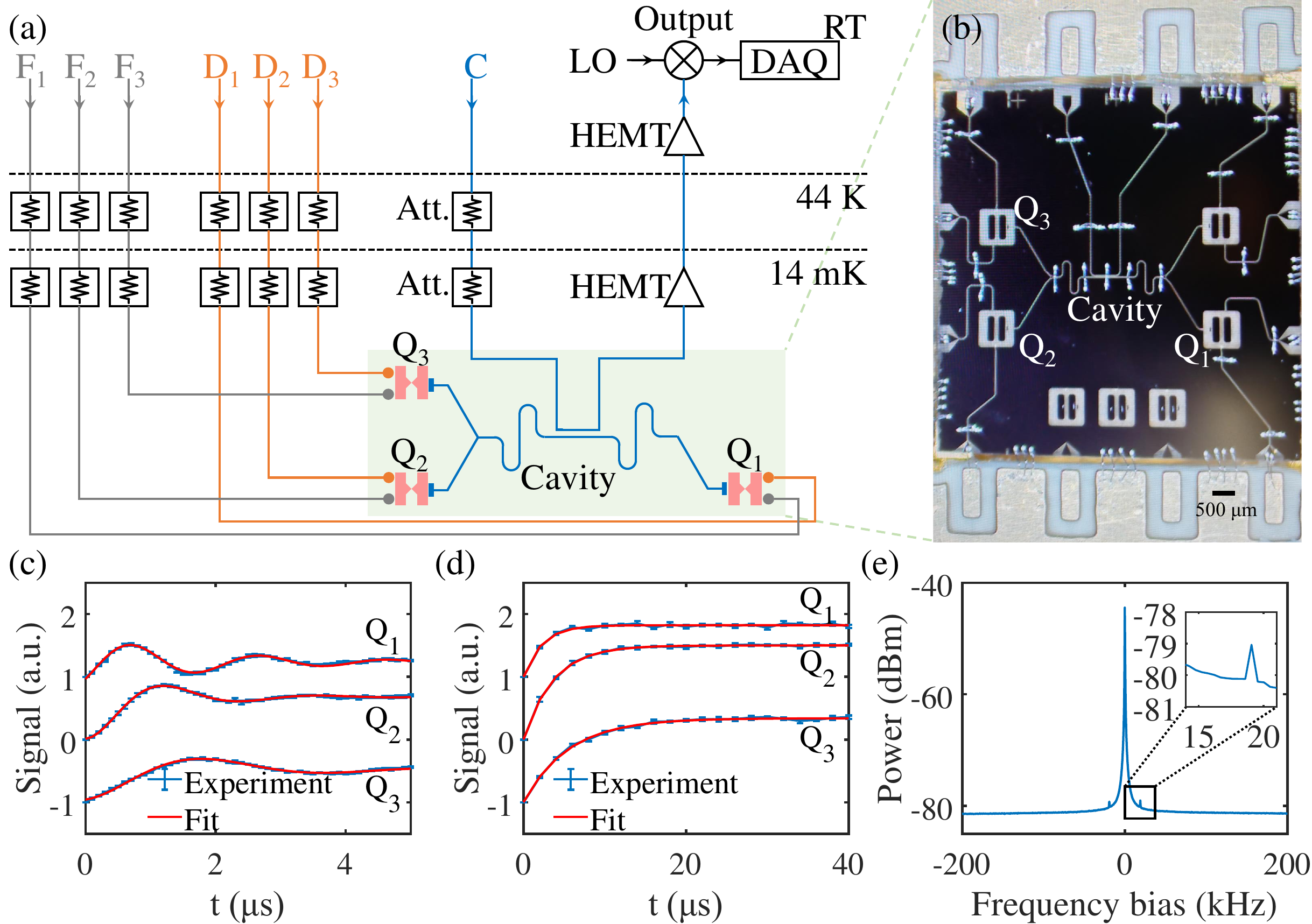}
	\caption{
		The experimental system.
		(a)	Schematic diagram of the experimental apparatus.
			It includes three flux lines ($\rm F_{1,2,3}$), three drive lines ($\rm D_{1,2,3}$), a cavity input line ($\rm C$), an output line, and a three-qubit chip (the green box).
		(b) Picture of the superconducting qubit chip.
		(c) Experimental results of Ramsey oscillations of the qubits.
		(d) Experimental results of $T_1$ measurements of the qubits.
		(e) Output spectrum with a simulated dark photon signal.
			}
	\label{Figure2}
\end{figure}

The heterodyne detection scheme, which is a powerful technology for measuring weak AC signals \cite{2022Wang1}, is applied to each qubit, as shown in Fig. \ref{Figure1}(d).
The Bloch vector of the qubit, driven by a resonant microwave, deviates from the North Pole of the Bloch sphere and reaches a new equilibrium position, as the solid light coral arrow in Fig. \ref{Figure1}(d) shows.
For readout, a pure-tone probe microwave of frequency $\omega_{\rm c}'$ is injected into the cavity, where $\omega_{\rm c}'$ is the frequency at which the absolute slope of the reflection coefficient curve of the cavity, $S_{11}$, is maximized.
The output of the cavity will also be a pure-tone signal with the power $P_{0}$.
When a dark-photon signal with frequency $\omega_{\rm X}$ is sensed by the qubit, the Bloch vector will oscillate around its equilibrium position at frequency $\delta = |\omega_{\rm X}-\omega_{\rm q}^{\rm (k)}|$ with amplitude $\varDelta\widehat{\tau}_{\rm z}^{\rm (k)}$.
Numerically, the amplitude is
\begin{equation}
  \begin{aligned}
	  \varDelta\widehat{\tau}_{\rm z}^{\rm (k)} = 1.5\times 10^{-4}\times\kappa(\frac{\chi}{10^{-15}})(\frac{\omega_{\rm X}/2\pi}{5\ {\rm GHz}})^{\frac{1}{2}}(\frac{C}{100\ {\rm fF}})\\
                                                (\frac{d}{300\ {\rm\mu m}})(\frac{T_{1}}{50\ {\rm\mu s}})(\frac{T_{2}^{*}}{100\ {\rm\mu s}})K(\delta),
  \end{aligned}
\end{equation}
where $d$ is the distance between the centers of the two conductor plates, $T_{1}$ and $T_{2}^{*}$ are the longitudinal and transverse relaxation time of the qubit, respectively, and $K$ is a unified line-shape function.
In such case, the resonant frequency $\omega_{\rm c}$ and reflectivity $S_{11}(\omega'_{\rm c})$ of the cavity will also oscillate at the frequency $\delta$, and consequently, two additional peaks with the power $P_{\pm}$ will appear at the sideband of the output of the cavity:
\begin{subequations}
	\begin{equation}
		P_{\pm} = \frac{\varDelta P^{2}}{16P_{0}},
    \end{equation}
    \begin{equation}
		\varDelta P = \varDelta\widehat{\tau}_{\rm z}^{\rm (k)}|\frac{\partial\omega_{\rm c}}{\partial\widehat{\tau}_{\rm z}^{\rm (k)}}||\frac{\partial S_{11}}{\partial\omega_{\rm c}}|\frac{P_{0}}{S_{11}(\omega_{\rm c}')}.
    \end{equation}
\end{subequations}
Their detunings from the main peak are both $\delta$.
It is noteworthy that, when the sideband peaks appeared, the dichotomy should be used to locate which qubit contributed the signals; while when a null result were obtained, constraints on the dark photons at frequencies $\omega_{\rm q}^{\rm (k)}\pm\delta$ for all values of $\rm k$ could be set simultaneously (see more details in Sec. \uppercase\expandafter{\romannumeral1} and \uppercase\expandafter{\romannumeral2} of Supplemental Material \cite{spp}).

A proof-of-principle experiment searching for dark photons was conducted.
The schematic diagram of the experimental apparatus is shown in Fig. \ref{Figure2}(a).
The entire setup was housed in a multistage cryogenic system.
A temperature as low as 14 mK was achieved in the last stage with a dilution refrigerator.
Seven input ports were employed. Port $\rm F_{k}$ was used to tune the frequency of the k-th qubit, Port $\rm D_{k}$ were used to drive the k-th qubits, and Port C was used to input the probe microwave to the cavity.
All the inputs were carefully attenuated so that the input noises were suppressed to a negligible level.
The signal reflected from the cavity was amplified by a cascade of high-electron-mobility transistor amplifiers (HEMTs).
The amplified signal was then mixed with a local oscillator signal, down-converted to 1 MHz, and finally acquired by a homemade data acquisition (DAQ) card based on FPGAs \cite{2020Tong,2024Kang}.
The three-qubit chip is highlighted by the green box in Fig. \ref{Figure2}(a) and more details are shown in Fig. \ref{Figure2}(b).
There were a 2-D cavity and three experimental qubits ($\rm Q_{1}$, $\rm Q_{2}$, and $\rm Q_{3}$) in the chip.
Each experimental qubit is connected to the cavity, a flux line, and a drive line.
When the strength of the magnetic flux was set to zero, the initial eigenfrequencies of the qubits were $\omega_{\rm q1} = 2\pi\times3780.686$ MHz, $\omega_{\rm q2} = 2\pi\times3831.086$ MHz, and $\omega_{\rm q3} = 2\pi\times3981.860$ MHz, respectively.
The capacitance $C$ and the distance $d$ between two conductor plates were 65 fF and 220 µm for all the qubits.
The package coefficient $\kappa$ was around 3.3, 0.9 and 0.3 for $\rm Q_{1}$, $\rm Q_{2}$, and $\rm Q_{3}$, respectively (see more details in Sec. \uppercase\expandafter{\romannumeral4} of Supplemental Material \cite{spp}).
The resonant frequency of the readout cavity is $\omega_{\rm c} = 2\pi\times 6595.211$ MHz.
Figure \ref{Figure2}(c) and (d) shows the Ramsey oscillation and the longitudinal relaxation of the qubits.
The transverse relaxation time $T_{2}^{*\rm Q_{1}} = 2.0\pm0.1$ µs, $T_{2}^{*\rm Q_{2}} = 1.1\pm0.1$ µs, $T_{2}^{*\rm Q_{3}} = 1.7\pm0.1$ µs, and longitudinal relaxation time $T_{1}^{\rm Q_{1}} = 2.3\pm0.1$ µs, $T_{1}^{\rm Q_{2}} = 3.7\pm0.1$ µs, $T_{1}^{\rm Q_{3}} = 5.5\pm0.2$ µs, are obtained by fitting the data.
The detection ability of the system was verified by injecting a pure-tone microwave with a frequency of 3780.669 MHz and a power of 0.1 mW.
The injected microwave was a simulation of a dark photon signal with the kinetic mixing $\chi^{\rm sim} = 6\times 10^{-12}$.
Figure \ref{Figure2}(e) shows the output spectrum of the cavity as the blue line.
Since the simulated dark photon signal was detuned from $\omega_{\rm q1}$ by $2\pi\times 19$ kHz, there were two sideband peaks at $\delta = \pm 2\pi\times19$ kHz.
The strength and positions of the sideband peaks are consistent with the expectation.

\begin{figure}[htp]
	\vspace{-2em}
	\centering
	\includegraphics[width=1\columnwidth]{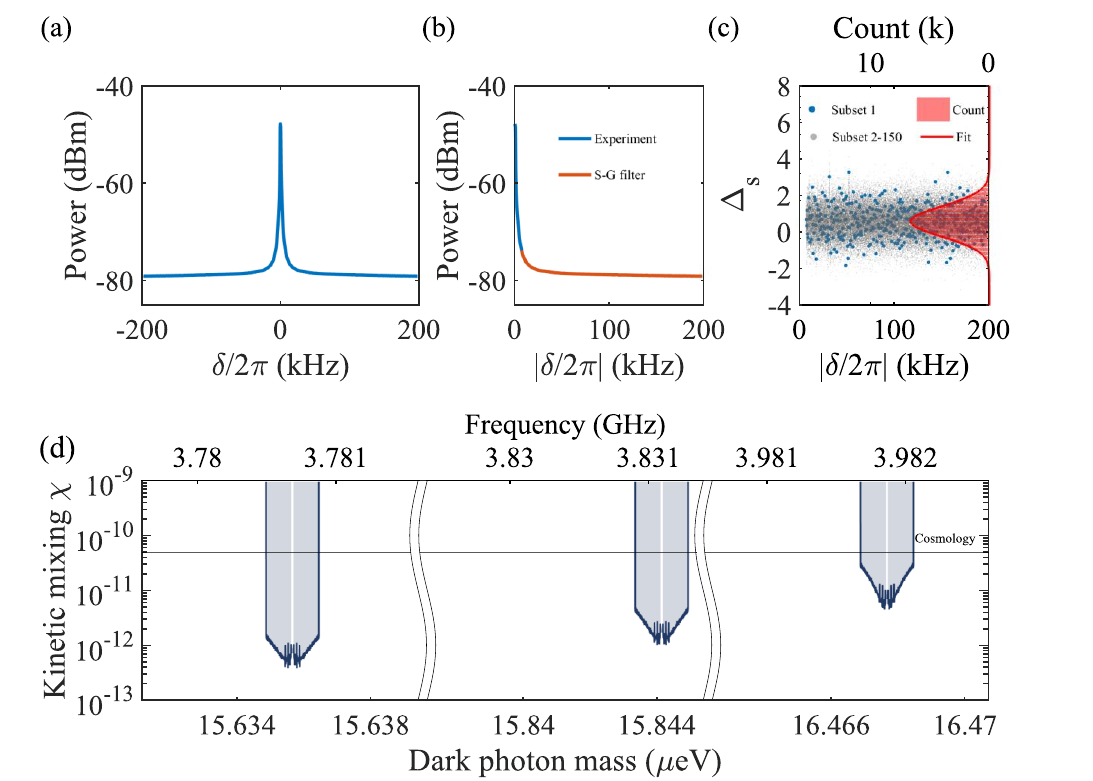}
	\caption{
		Results from the first dataset of the experiment.
		(a)	Two-sided spectrum of Subset 1.
		(b) Single-sided spectrum of Subset 1 and the S-G filter.
		(c) Standardized power residuals and their distribution.
		(d) The final constraints on dark photons set by this work (dark blue) and the cosmology constraints (black) \cite{2012Arias}.
			}
	\label{Figure3}
\end{figure}

The dark photon searching experiment is performed without the simulated dark photon signal.
The first dataset contains 60000 raw spectra, corresponding to an integration time of 9 hours.
These raw spectra are divided into 150 subsets, each containing 400 raw spectra.
Each subset of raw spectra is averaged to suppress the noises.
Figure \ref{Figure3}(a) shows the average spectrum of Subset 1.
Given that the dark photon signals are expected to be symmetric around $\delta = 0$, the two-sided spectrum is converted to a single sided spectrum in order to improve the signal-to-noise-ratio, as Fig. \ref{Figure3}(b) shows.
Since only the fluctuations rather than the noise itself are concerned, the Savitzky-Golay (SG) filter is applied to remove the baseline of the power spectrum, as the orange lines in Fig \ref{Figure3}(b) shows \cite{1964Savitzky}.
Thirty points with strong backgrounds at the center were discarded because the SG filter failed here.
The blue dots in Fig. \ref{Figure3}(c) depict the standardized power residuals $\Delta_{\rm s}$ of Subset 1,
\begin{equation}
	\Delta_{\rm s} = \frac{P-P^{\rm SG}}{\sigma},
\end{equation}
where $P$ is the noise power, $P^{\rm SG}$ is the S-G filter function, and $\sigma$ is the standard deviation of the noise power.
The gray dots depict the standardized power residuals of Subset 2-150.
A standardized power residual exceeding 5 is referred to as a candidate signal.
The distribution of the standardized power residuals is illustrated by the red bars in Fig. \ref{Figure3}(c), and the Gaussian fitting result is shown as the red line.
No standardized power residual over 5 is observed.
Therefore, an exclusion of dark photons can be obtained from experimental results.
The data analysis method that gives the upper bound on the kinetic mixing is similar to that developed by ADMX \cite{2024Kang,2022Cervantes2} (see more details in Sec. \uppercase\expandafter{\romannumeral5} of Supplemental Material \cite{spp}).
Figure \ref{Figure3}(d) shows the upper bounds on the kinetic mixing with a confidence level of 90$\%$.
The left, middle and right parts of Fig. \ref{Figure3}(d) refer to the constraints obtained by considering $\rm Q_{1}$, $\rm Q_{2}$, and $\rm Q_{3}$, respectively.
An upper bound of $\chi < 4\times10^{-13}$ is achieved at 3780.718 MHz, corresponding to the dark photon mass of 15.636 µeV.
This result exceeds the constraints from CMB distortions \cite{2012Arias} by more than two orders of magnitude.

\begin{figure}[http]
	\centering
	\includegraphics[width=1\columnwidth]{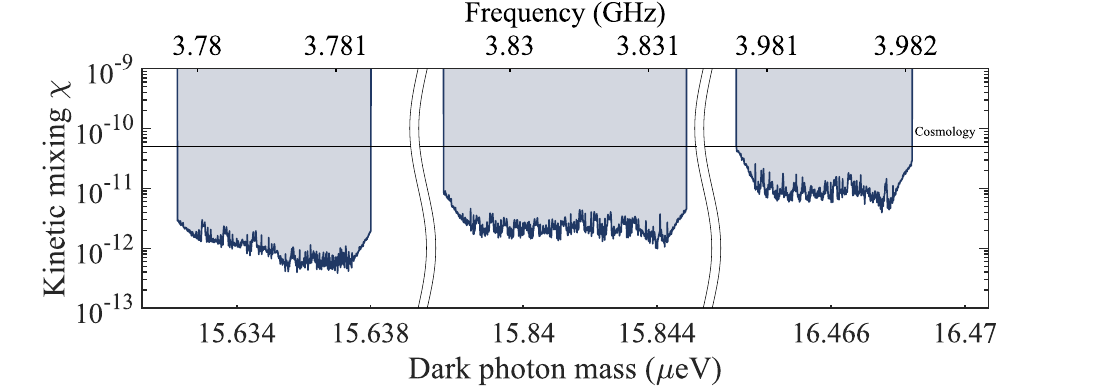}
	\caption{
		Combined results from all data sets of the experiment.
			}
	\label{Figure4}
\end{figure}

In order to explore more parameter space of dark photons, the eigenfrequencies of the qubits were tuned through a 1 MHz range.
The external magnetic flux was generated by injecting DC current from Port $\rm F_{k}$.
The eigenfrequency of $\rm Q_{1}$, $\rm Q_{2}$, and $\rm Q_{3}$ swept through the range $\rm 3780.041\ MHz\sim 3781.062\ MHz$, $\rm 3829.710\ MHz\sim 3831.086\ MHz$, $\rm 3980.967\ MHz\sim 3981.860\ MHz$, respectively, with the current changing from 0 µA to 29 µA in 17 steps (see more details in Sec. \uppercase\expandafter{\romannumeral4} of Supplemental Material \cite{spp}).
Each tuning was followed by 9-hour data collection, resulting in 17 datasets.
The parameter space excluded by all datasets is plotted as the blue regions in Fig. \ref{Figure4}.
The results demonstrate the feasibility and practicability of the method proposed in this article.

\begin{table}[htp]
    \centering
    \renewcommand\arraystretch{1}
    \begin{tabular}{c | c c c c c c}
        \hline
        {Config.}              &{\R{1}}&{\R{2}}&{\R{3}}&{\R{4}}&{\R{5}}&{\R{6}}\\
        \cline{1-7}
        {$\omega_{\rm q}/2\pi$ (GHz)}         &{3$\sim$5}&{5$\sim$10}&{10$\sim$15}&{15$\sim$20}&{20$\sim$25}&{25$\sim$30}\\
		    {$\varpi_{\rm q}/2\pi$ (MHz)}         &{0.26    }&{0.26     }&{0.25      }&{0.27      }&{0.37      }&{0.60}\\
		    {$\varDelta\omega_{\rm q}/2\pi$ (MHz)}&{5.2     }&{5.2      }&{5.0       }&{5.4       }&{7.4       }&{12}\\
        \hline
    \end{tabular}
    \caption{Parameters of the detection system.}
    \label{Table1}
\end{table}

\begin{figure}[http]
	\centering
	\includegraphics[width=1\columnwidth]{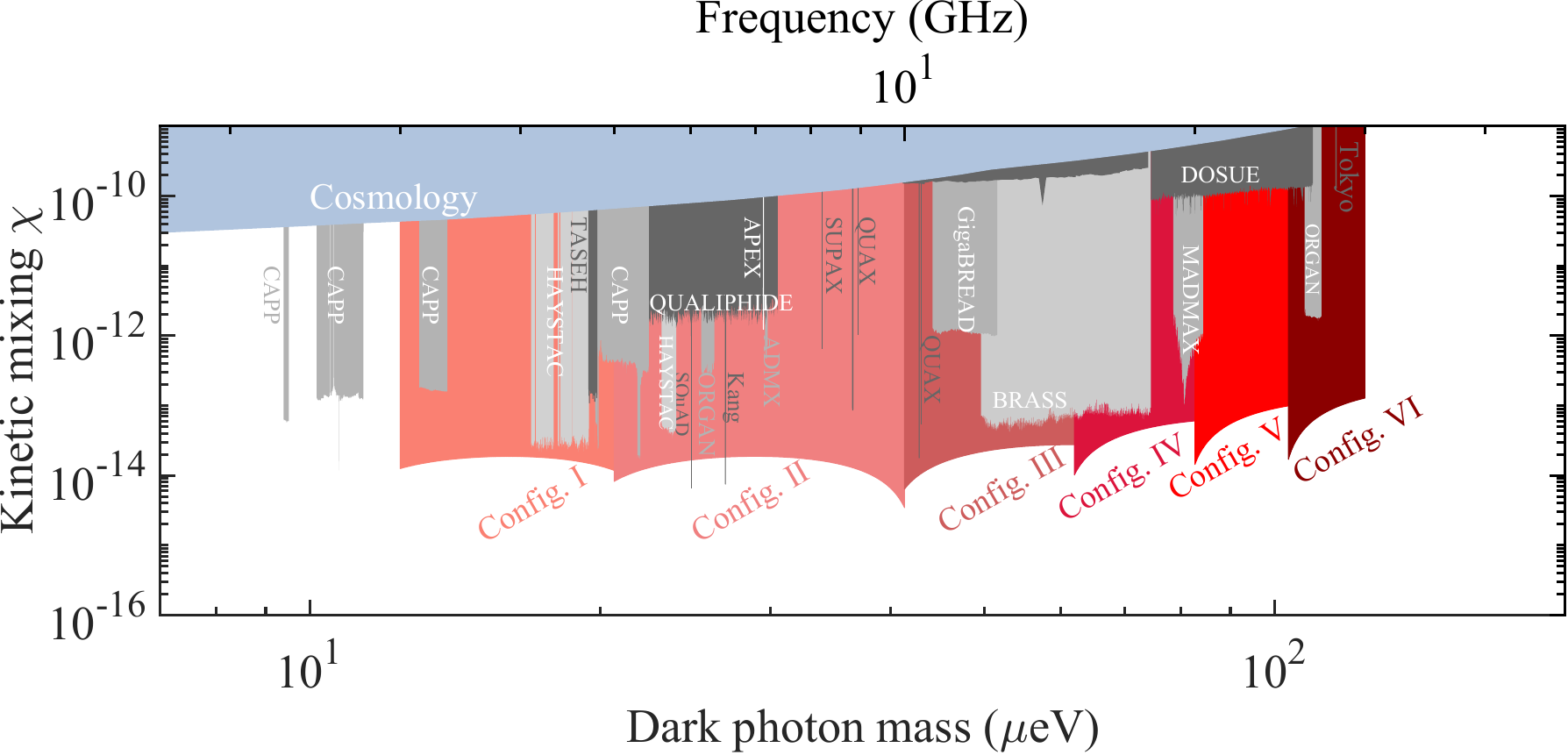}
	\caption{
		The expected exclusion limits on the kinetic mixing.
		The salmon, light coral, Indian red, crimson, red and dark red regions refer to the expectation of Config. \R{1}, \R{2}, \R{3}, \R{4}, \R{5}, and \R{6}, respectively.
		For each configuration, the integration time is set to twenty days.
		The light steel blue region stands for the cosmological constraints and the gray regions refer to previous haloscope constraints.
		Data are adapted from \cite{2020Hare}.
			}
	\label{Figure5}
\end{figure}

Our work opens up the door to scaling up to numerous superconducting qubits on a single chip for scalably searching for dark photons.
Herein we focus on six configurations with qubits evenly distributed the frequency ranges of 3$\sim$5 GHz (Config. \R{1}), 5$\sim$10 GHz (Config. \R{2}), 10$\sim$15 GHz (Config. \R{3}), 15$\sim$20 GHz (Config. \R{4}), 20$\sim$25 GHz (Config. \R{5}), and 25$\sim$30 GHz (Config. \R{6}).
The frequency interval between neighboring qubits $\varDelta\omega_{\rm q}$ is set to 20 times of the effective sensitive width $\varpi_{\rm q}$, so that the targeted frequency range can be completely covered within 20 times of tuning, as listed in Table \ref{Table1} (see more details in Sec. \uppercase\expandafter{\romannumeral3} of Supplemental Material \cite{spp}).
For each configuration, the resonant frequency of the readout cavity $\omega_{\rm c}$ is set to be 0.5 GHz lower than the lower bound of the qubit frequency range.
Although 2-D cavities with quality factors of several million and superconducting qubits with relaxation time of several milliseconds have been experimentally achieved, here we choose the conservative values $Q_{\rm c} = 2\times 10^{4}$, $T_{1} = 50$ µs and $T_{2}^{*} = 100$ µs considering the yield \cite{2012Megrant,2022Shi,2022Lozano,2021Place,2022Wang2,2024Ganjam}.
The capacitance $C$ and the distance between two conductor plates $d$ are set to be 100 fF and 300 µm, respectively.
The coupling between each qubit and the readout cavity is expected to be $2\pi\times 100$ MHz.
The physical temperature of the entire experimental setup $T_{\rm phys}$ is assumed to be 10 mK, which can be achieved by commercial dilution refrigerators.
The maximum probe power of the cavity is decided by the critical photon number $n_{\rm c} = \min\left\{\Omega_{\rm c}^{\rm(k)2}/4G^{2}\right\}$ \cite{2004Blais}.
The metallic surroundings of the setup is chosen to be a hollow cylinder with the radius $R$, where $J_{0}(\omega_{\rm c}R/c) = 0$, and $J_{0}$ is the Bessel function of the first kind.
As a result, the package coefficient $\kappa$ is $1-J_{0}(0)/J_{0}(\omega_{\rm q}R/c)$.
The expected constraints on dark photons from Config. 1$\sim$6 are shown as the red regions in Fig. \ref{Figure5} with different shades.
In all configurations, the integration time is 1 day for every tuning and thus 20 days for the total frequency range.
An improvement over the cosmological results (the light steel blue region) of 3$\sim$5 orders of magnitude is achievable.
The gray regions in Fig. \ref{Figure5} stand for previous haloscope results.

In conclusion, we have proposed a scalable scheme for searching for dark photons with a superconducting qubit architecture.
The scheme takes advantage of the remarkable scalability of superconducting qubits as well as their sensitive response to the dark-photon-induced electric field.
The scalable superconducting qubit architecture can enable searching for dark matter with high sensitivity in a wide mass range.
A proof-of-principle experiment has been successfully realized with a three-qubit chip, and it established the most stringent constraints on dark photons around 15.636 µeV, 15.844 µeV, and 16.467 µeV, exceeding the previous results by 1$\sim$2 orders of magnitude.
It is anticipated that, with delicately designed setups, stringent constraints on the kinetic mixing of dark photons can be set mass range of $12\sim 66$ µeV.
Furthermore, our scheme can also be adapted to search for axions and axion-like particles with an external magnetic field (see more details in Sec. \uppercase\expandafter{\romannumeral6} of Supplemental Material \cite{spp}).
This work unveils the tremendous application potential of superconducting qubits in the field of precision measurement, and sets up an unprecedented path for scalable, high-performance on-chip dark matter searching in future.

This work was supported by the Innovation Program for Quantum Science and Technology (2021ZD0302200), the Chinese Academy of Sciences (No. GJJSTD20200001), the National Key R\&D Program of China (Grants No. 2021YFB3202800, No. 2018YFA0306600, No. 2021YFC2203100), Anhui Initiative in Quantum Information Technologies (Grant No. AHY050000), NSFC (12150010, 12205290, 12261160569, 12261131497).  Y. F. C. and M. J.  thank the Fundamental Research Funds for Central Universities. Y.F.C. is supported in part by CAS Young Interdisciplinary Innovation Team (JCTD-2022-20), 111 Project (B23042). M. J. is supported in part by China Postdoctoral Science Foundation (2022TQ0330). This work was partially carried out at the USTC Center for Micro and Nanoscale Research and Fabrication.

\appendix

\section{Principle of heterodyne detection}

For simplicity, the Hamiltonian of the cavity is neglected in this section.
The Hamiltonian of a qubit is
\begin{equation}
  \begin{aligned}
    \mathcal{H} = \frac{1}{2}\hbar\omega_{\rm q}\widehat{\tau}_{\rm z} + \hbar\nu(\widehat{\tau}_{+}e^{-i\omega_{\rm d}t}+\widehat{\tau}_{-}e^{i\omega_{\rm d}t})\\
                  + \hbar\eta(\widehat{\tau}_{+}e^{-i(\omega_{\rm X}t+\phi)}+\widehat{\tau}_{-}e^{i(\omega_{\rm X}t+\phi)}),
  \end{aligned}
  \label{Eq1}
\end{equation}
where $\widehat{\tau}_{\rm z,\pm}$ are the Pauli operators of the qubit, $\omega_{\rm q}$ is the eigenfrequency of the qubit, $\nu$ is the driving strength, $\eta$ and $\phi$ are the strength and phase of the dark photon field, respectively.
The dark photon field strength $\eta$ can be expressed as
\begin{equation}
  \begin{aligned}
    \eta = \frac{1}{2\sqrt{2}}E_{\rm X}d\sqrt{C\omega_{\rm X}/\hbar},
  \end{aligned}
  \label{Eq2}
\end{equation}
where $E_{\rm X}$ is the electric field induced by dark matter, $C$ is the capacitance of the qubit, $d$ is the distance between two conductor plates, and $\omega_{\rm X}$ is the frequency of the dark matter field.
The Heisenberg equations of motion are
\begin{subequations}
  \begin{equation}
    \begin{aligned}
      \dot{\widehat{\tau}}_{\rm z} = &-i2\nu(\widehat{\tau}_{+}e^{-i\omega_{\rm d}t}-\widehat{\tau}_{-}e^{i\omega_{\rm d}t}) \\
                                     &-i2\eta(\widehat{\tau}_{+}e^{-i(\omega_{\rm X}t+\phi)}-\widehat{\tau}_{-}e^{i(\omega_{\rm X}t+\phi)}) - \frac{\widehat{\tau}_{\rm z}-\widehat{\tau}_{\rm z}^{0}}{T_{1}},\\
    \end{aligned}
    \label{Eq3a}
  \end{equation}
  \begin{equation}
    \dot{\widehat{\tau}}_{-}       = -i\omega_{\rm q}\widehat{\tau}_{-} + i\nu\widehat{\tau}_{\rm z}e^{-i\omega_{\rm d}t} + i\eta\widehat{\tau}_{\rm z}e^{-i(\omega_{\rm X}t+\phi)} - \frac{\widehat{\tau}_{-}}{T_{2}^{*}},
    \label{Eq3b}
  \end{equation}
\end{subequations}
where $\widehat{\tau}_{\rm z}^{0} = -1$ for a thermally polarized two-level system.
It is natural to assume that the solution takes the following forms:
\begin{subequations}
  \begin{equation}
    \widehat{\tau}_{-}     = \mathcal{M}e^{-i\omega_{\rm d}t} + \mathcal{N}e^{-i(\omega_{\rm X}t+\phi)},
    \label{Eq4a}
  \end{equation}
  \begin{equation}
    \widehat{\tau}_{\rm z} = \gamma + \alpha\cos(\delta t+\phi) + \beta\sin(\delta t+\phi),\\
    \label{Eq4b}
  \end{equation}
\end{subequations}
where $\delta = |\omega_{\rm X} - \omega_{\rm d}|$, then
\begin{subequations}
  \begin{equation}
    \dot{\widehat{\tau}}_{-}     = -i\omega_{\rm d}\mathcal{M}e^{-i\omega_{\rm d}t} -i\omega_{\rm X}\mathcal{N}e^{-i(\omega_{\rm X}t+\phi)},
    \label{Eq5a}
  \end{equation}
  \begin{equation}
    \dot{\widehat{\tau}}_{\rm z} = -\delta\alpha\sin(\delta t+\phi) + \delta\beta\cos(\delta t+\phi).
    \label{Eq5b}
  \end{equation}
\end{subequations}
By substituting Eq. \ref{Eq4a},\ref{Eq4b}, and \ref{Eq5a} for $\widehat{\tau}_{-}$, $\widehat{\tau}_{\rm z}$, and $\dot{\widehat{\tau}}_{-}$ in Eq. \ref{Eq3b}, respectively, we obtain
\begin{subequations}
  \begin{equation}
    \mathcal{M} = \frac{i\nu \gamma + i\eta\alpha/2+ \eta\beta/2}{-i\Omega_{\rm d}+1/T_{2}^{*}},
    \label{Eq6a}
  \end{equation}
  \begin{equation}
    \mathcal{N} = \frac{i\eta\gamma+ i\nu \alpha/2 - \nu \beta/2}{-i\Omega_{\rm X}+1/T_{2}^{*}},
    \label{Eq6b}
  \end{equation}
  \begin{equation}
    \widehat{\tau}_{-} = \frac{i\nu \gamma + i\eta\alpha/2+ \eta\beta/2}{-i\Omega_{\rm d}+1/T_{2}^{*}}e^{-i\omega_{\rm d}t} + \frac{i\eta\gamma+ i\nu \alpha/2 - \nu \beta/2}{-i\Omega_{\rm X}+1/T_{2}^{*}}e^{-i(\omega_{\rm X}t+\phi)},
    \label{Eq6c}
  \end{equation}
\end{subequations}
where $\Omega_{\rm d,X} = \omega_{\rm d,X} - \omega_{\rm q}$.
Note that when the qubit is resonantly driven, $\Omega_{\rm d} = 0$ and $\delta = |\omega_{\rm X} - \omega_{\rm q}|$.

Then by substituting Eq. \ref{Eq6c} for $\widehat{\tau}_{-}$ in Eq. \ref{Eq3a}, we get
\begin{equation}
  \begin{aligned}
    \dot{\widehat{\tau}}_{\rm z} = &-\delta\alpha\sin(\delta t+\phi) + \delta\beta\cos(\delta t+\phi)\\
                                   = &-2\nu [\frac{(2\nu \gamma+\eta\alpha)/T_{2}^{*}+\eta\beta\Omega_{\rm d}}{\Omega_{\rm d}^{2}+1/T_{2}^{*2}}\\
                                     & + \frac{(2\eta\gamma+\nu \alpha)/T_{2}^{*}-\nu \beta\Omega_{\rm X}}{\Omega_{\rm X}^{2}+1/T_{2}^{*2}}\cos(\delta t+\phi)\\
                                     &+ \frac{(2\eta\gamma+\nu \alpha)\Omega_{\rm X}+\nu \beta/T_{2}^{*}}{\Omega_{\rm X}^{2}+1/T_{2}^{*2}}\sin(\delta t+\phi)]\\
                                     &-2\eta[\frac{(2\eta\gamma+\nu \alpha)/T_{2}^{*}-\nu \beta\Omega_{\rm X}}{\Omega_{\rm X}^{2}+1/T_{2}^{*2}}\\
                                     &+ \frac{(2\nu \gamma+\eta\alpha)/T_{2}^{*}+\eta\beta\Omega_{\rm d}}{\Omega_{\rm d}^{2}+1/T_{2}^{*2}}\cos(\delta t+\phi)\\
                                     &- \frac{(2\nu \gamma+\eta\alpha)\Omega_{\rm d}-\eta\beta/T_{2}^{*}}{\Omega_{\rm d}^{2}+1/T_{2}^{*2}}\sin(\delta t+\phi)]\\
                                     &-[\gamma + \alpha\cos(\delta t+\phi) + \beta\sin(\delta t+\phi)]/T_{1} + \widehat{\tau}_{\rm z}^{0}/T_{1}.
  \end{aligned}
  \label{Eq7}
\end{equation}
This equation is satisfied if and only if the coefficients of $1, \cos(\delta t+\phi)$ and $\sin(\delta t+\phi)$ cancel respectively, i.e.,
\begin{widetext}
\begin{equation}
  \begin{bmatrix}
    &\frac{1}{T_{1}} + \frac{4\nu^2/T_{2}^{*}}{\Omega_{\rm d}^{2}+1/T_{2}^{*2}}+\frac{4\eta^2/T_{2}^{*}}{\Omega_{\rm X}^{2}+1/T_{2}^{*2}}   &\frac{2\nu\eta/T_{2}^{*}}{\Omega_{\rm d}^{2}+1/T_{2}^{*2}}+\frac{2\nu\eta/T_{2}^{*}}{\Omega_{\rm X}^{2}+1/T_{2}^{*2}}                          &\frac{2\nu\eta\Omega_{\rm d}}{\Omega_{\rm d}^{2}+1/T_{2}^{*2}}-\frac{2\nu\eta\Omega_{\rm X}}{\Omega_{\rm X}^{2}+1/T_{2}^{*2}}\\
    &\frac{2\nu\eta/T_{2}^{*}}{\Omega_{\rm d}^{2}+1/T_{2}^{*2}}+\frac{2\nu\eta/T_{2}^{*}}{\Omega_{\rm X}^{2}+1/T_{2}^{*2}}                  &\frac{1}{2T_{1}} + \frac{\nu^2/T_{2}^{*}}{\Omega_{\rm X}^{2}+1/T_{2}^{*2}}+\frac{\eta^2/T_{2}^{*}}{\Omega_{\rm d}^{2}+1/T_{2}^{*2}}            &\frac{\delta}{2} - \frac{\nu^2\Omega_{\rm X}}{\Omega_{\rm X}^{2}+1/T_{2}^{*2}}+\frac{\eta^2\Omega_{\rm d}}{\Omega_{\rm d}^{2}+1/T_{2}^{*2}}\\
    &-\frac{2\nu\eta\Omega_{\rm d}}{\Omega_{\rm d}^{2}+1/T_{2}^{*2}}+\frac{2\nu\eta\Omega_{\rm X}}{\Omega_{\rm X}^{2}+1/T_{2}^{*2}}         &-\frac{\delta}{2} + \frac{\nu^2\Omega_{\rm X}}{\Omega_{\rm X}^{2}+1/T_{2}^{*2}}-\frac{\eta^2\Omega_{\rm d}}{\Omega_{\rm d}^{2}+1/T_{2}^{*2}}   &\frac{1}{2T_{1}} + \frac{\nu^2/T_{2}^{*}}{\Omega_{\rm X}^{2}+1/T_{2}^{*2}}+\frac{\eta^2/T_{2}^{*}}{\Omega_{\rm d}^{2}+1/T_{2}^{*2}}
  \end{bmatrix}
  \begin{bmatrix}  \gamma \\ \alpha \\ \beta \end{bmatrix} =
  \begin{bmatrix}  \widehat{\tau}_{\rm z}^{0}/T_{1}\\ 0 \\ 0 \end{bmatrix}.\ \
  \label{Eq8}
\end{equation}
Since the dark photon field is expected to be extremely weak, i.e. $\nu^{2}>>\nu\eta>>\eta^{2}$ and $\gamma>>\alpha,\beta$, Eq. \ref{Eq8} can be simplified to
\begin{equation}
  \begin{bmatrix}
    &\frac{1}{T_{1}} + \frac{4\nu^2/T_{2}^{*}}{\Omega_{\rm d}^{2}+1/T_{2}^{*2}}                                                       &0                                                                                 &0\\
    &\frac{2\nu\eta/T_{2}^{*}}{\Omega_{\rm d}^{2}+1/T_{2}^{*2}}+\frac{2\nu\eta/T_{2}^{*}}{\Omega_{\rm X}^{2}+1/T_{2}^{*2}}            &\frac{1}{2T_{1}} + \frac{\nu^2/T_{2}^{*}}{\Omega_{\rm X}^{2}+1/T_{2}^{*2}}        &\frac{\delta}{2} - \frac{\nu^2\Omega_{\rm X}}{\Omega_{\rm X}^{2}+1/T_{2}^{*2}}\\
    &-\frac{2\nu\eta\Omega_{\rm d}}{\Omega_{\rm d}^{2}+1/T_{2}^{*2}}+\frac{2\nu\eta\Omega_{\rm X}}{\Omega_{\rm X}^{2}+1/T_{2}^{*2}}   &-\frac{\delta}{2} + \frac{\nu^2\Omega_{\rm X}}{\Omega_{\rm X}^{2}+1/T_{2}^{*2}}   &\frac{1}{2T_{1}} + \frac{\nu^2/T_{2}^{*}}{\Omega_{\rm X}^{2}+1/T_{2}^{*2}}
  \end{bmatrix}
  \begin{bmatrix}  \gamma \\ \alpha \\ \beta \end{bmatrix} =
  \begin{bmatrix}  \widehat{\tau}_{\rm z}^{0}/T_{1}\\ 0 \\ 0 \end{bmatrix}.
  \label{Eq9}
\end{equation}
Consequently, the solutions are
\begin{subequations}
  \begin{equation}
    \gamma = \widehat{\tau}_{\rm z}^{0}\frac{1}{1+\frac{4\nu^2T_{1}T_{2}^{*}}{1+\Omega_{\rm d}^{2}T_{2}^{*2}}},
    \label{Eq10a}
  \end{equation}
  \begin{equation}
    \alpha = -2\nu\eta\gamma\frac{(\frac{\delta}{2} - \frac{\nu^2\Omega_{\rm X}}{\Omega_{\rm X}^{2}+1/T_{2}^{*2}})(\frac{\Omega_{\rm d}}{\Omega_{\rm d}^{2}+1/T_{2}^{*2}}-\frac{\Omega_{\rm X}}{\Omega_{\rm X}^{2}+1/T_{2}^{*2}})+(\frac{1}{2T_{1}} + \frac{\nu^2/T_{2}^{*}}{\Omega_{\rm X}^{2}+1/T_{2}^{*2}})(\frac{1/T_{2}^{*}}{\Omega_{\rm d}^{2}+1/T_{2}^{*2}}+\frac{1/T_{2}^{*}}{\Omega_{\rm X}^{2}+1/T_{2}^{*2}})}{(\frac{\delta}{2} - \frac{\nu^2\Omega_{\rm X}}{\Omega_{\rm X}^{2}+1/T_{2}^{*2}})^2 + (\frac{1}{2T_{1}} + \frac{\nu^2/T_{2}^{*}}{\Omega_{\rm X}^{2}+1/T_{2}^{*2}})^2},
    \label{Eq10b}
  \end{equation}
  \begin{equation}
    \beta  = -2\nu\eta\gamma\frac{(\frac{\delta}{2} - \frac{\nu^2\Omega_{\rm X}}{\Omega_{\rm X}^{2}+1/T_{2}^{*2}})(\frac{1/T_{2}^{*}}{\Omega_{\rm d}^{2}+1/T_{2}^{*2}}+\frac{1/T_{2}^{*}}{\Omega_{\rm X}^{2}+1/T_{2}^{*2}})-(\frac{1}{2T_{1}} + \frac{\nu^2/T_{2}^{*}}{\Omega_{\rm X}^{2}+1/T_{2}^{*2}})(\frac{\Omega_{\rm d}}{\Omega_{\rm d}^{2}+1/T_{2}^{*2}}-\frac{\Omega_{\rm X}}{\Omega_{\rm X}^{2}+1/T_{2}^{*2}})}{(\frac{\delta}{2} - \frac{\nu^2\Omega_{\rm X}}{\Omega_{\rm X}^{2}+1/T_{2}^{*2}})^2 + (\frac{1}{2T_{1}} + \frac{\nu^2/T_{2}^{*}}{\Omega_{\rm X}^{2}+1/T_{2}^{*2}})^2}.
    \label{Eq10c}
  \end{equation}
\end{subequations}
The amplitude of the oscillation of $\widehat{\tau}_{\rm z}$ is $\varDelta\widehat{\tau}_{\rm z} = \sqrt{\alpha^2+\beta^2}$.
\end{widetext}

\section{Output power of the cavity}

As the readout cavity is pumped by a pure tone microwave with the frequency $\omega_{\rm c}'$ and the power $P_{\rm in}$, the output power is
\begin{equation}
    P_{0} = P_{\rm in}S_{11}(\omega_{\rm c}'),
    \label{Eq11}
\end{equation}
where $S_{11}$ is the reflection coefficient of the cavity.
A general expression of the reflection coefficient is
\begin{equation}
    S_{11} = |1-\frac{\theta}{1+2iQ_{\rm L}(\omega/\omega_{\rm c}-1)}|^{2},
    \label{Eq12}
\end{equation}
where $\theta$ is a variant related to the coupling strength $\beta$, and $Q_{\rm L}$ is the loaded quality factor.
For 2-D and 3-D cavities, $\theta$ is $\beta/(1+\beta)$ and $2\beta/(1+\beta)$, respectively \cite{2015Probst,2020Kudra}.

When the qubit is driven by a signal and thus oscillates with the frequency $\delta$ and the amplitude $\Delta\widehat{\tau}_{\rm z}$, the output power of the cavity will also oscillate at the same frequency,
The amplitude is
\begin{equation}
    \varDelta P = \varDelta\widehat{\tau}_{\rm z}^{\rm (k)}|\frac{\partial\omega_{\rm c}}{\partial\widehat{\tau}_{\rm z}}||\frac{\partial S_{11}}{\partial\omega_{\rm c}}|P_{\rm in},
    \label{Eq13}
\end{equation}
where $|\partial\omega_{\rm c}/\partial\widehat{\tau}_{\rm z}| = g$, and $g = G^{2}/\Omega$.
To optimize the signal-to-noise ratio (SNR), as large $|\partial S_{11}/\partial\omega_{\rm c}|$ as possible is desired.
According to Eq. \ref{Eq12}, $|\partial S_{11}/\partial\omega_{\rm c}|$ is maximized as $\omega_{\rm c}' = \omega_{\rm c}(1\pm1/2\sqrt{3}Q_{\rm L})$.

The oscillation of a pure-tone microwave can result in two sideband peaks in the frequency domain.
Their frequencies are $\omega_{\pm} = \omega_{\rm c}'\pm\delta$.
Since the electric component of the output signal is $E = \sqrt{P}\exp(i\omega_{\rm c}'t)$ and $P = P_{0} + \varDelta P\cos(\delta t+\psi)$,
the first-order approximation of the electric component is
\begin{equation}
      E = \sqrt{P_{0}}e^{i\omega_{\rm c}'t} + \frac{\varDelta P}{4\sqrt{P_{0}}}e^{i(\omega_{+}t+\psi)} + \frac{\varDelta P}{4\sqrt{P_{0}}}e^{i(\omega_{-}t-\psi)},
      \label{Eq14}
\end{equation}
resulting in the sideband peak power
\begin{equation}
      P_{\pm} = \frac{27}{16}P_{0}|\zeta|^{2}|\varDelta\widehat{\tau}_{\rm z}^{\rm (k)}g\frac{Q_{\rm L}}{\omega_{\rm c}}|^{2},
      \label{Eq15}
\end{equation}
where $\zeta = (2\beta+\beta^2)/(4+2\beta+\beta^2)$ for 2-D cavities and $\zeta = \beta/(1-\beta+\beta^{2})$ for 3-D cavities.

\section{Noise analysis}

\subsection{Projection noise}
For the k-th qubit, the sensitivity is optimized when $\gamma = \widehat{\tau}_{\rm z}^{0(k)}/2$, i.e., the populations of the qubit at the ground state and the excited state are
\begin{subequations}
  \begin{equation}
      p_{\rm g}^{\rm (k)} = 0.75,
  \label{Eq16a}
  \end{equation}
  \begin{equation}
      p_{\rm e}^{\rm (k)} = 0.25,
  \label{Eq16b}
  \end{equation}
\end{subequations}
respectively.
Therefore, the fluctuation of $\widehat{\tau}_{\rm z}^{\rm (k)}$ is
\begin{equation}
  \widehat{\varsigma}_{\rm z}^{\rm(k)}(\omega) = 2\sqrt{p_{\rm g}^{\rm (k)}p_{\rm e}^{\rm (k)}}S(\omega,0;T_{1}),
  \label{Eq17}
\end{equation}
where $S(\omega)$ is the Lorentzian spectral density.
According to Eq. \ref{Eq13}, this fluctuation of $\widehat{\tau}_{\rm z}^{\rm (k)}$ can induce a fluctuation in the output power of the cavity,
\begin{subequations}
  \begin{equation}
    P_{\rm proj} = \sum_{k}P_{\rm proj}^{\rm(k)},
    \label{Eq18a}
  \end{equation}
  \begin{equation}
    P_{\rm proj}^{\rm(k)} = \widehat{\varsigma}_{\rm z}^{\rm(k)}|\frac{\partial\omega_{\rm c}}{\partial\widehat{\tau}_{\rm z}}||\frac{\partial S_{11}}{\partial\omega_{\rm c}}|P_{\rm in}.
    \label{Eq18b}
  \end{equation}
\end{subequations}

\subsection{Black-body radiation (BR) noise and zero-point fluctuation (zpf) noise}
The energy density due to black-body radiation is
\begin{equation}
  u(\omega)d\omega = \frac{\hbar\omega^{3}}{\pi^{2}c^{3}}\frac{1}{e^{\hbar\omega/k_{\rm B}T_{\rm phys}}-1} d\omega.
  \label{Eq19}
\end{equation}
The noise is contributed by the electrical component of the black-body radiation,
\begin{equation}
  E_{\rm BR}(\omega) = \sqrt{\frac{u(\omega)B}{2\epsilon_{0}}},
  \label{Eq20}
\end{equation}
where $\epsilon_{0}$ is the vacuum permittivity,  $B$ is the frequency bin of the spectrum, and the factor 2 comes from the equipartition theorem.
The black-body radiation is usually accompanied by zero-point fluctuations.
Since the metallic surroundings around the experimental setup form a boundary condition, which can be considered as a cavity with the quality factor $Q'$ and resonant frequencies $\omega'$, the zero-point energy is limited around these resonant frequencies with the total energy of $\hbar\omega'/2$.
The zpf electric field sensed by a single qubit is
\begin{equation}
  E_{\rm zpf}(\omega) = \sqrt{\frac{\frac{1}{2}\hbar\omega'}{2\epsilon_{0}V'}S(\omega,\omega';Q'/\omega')B},
  \label{Eq21}
\end{equation}
where $V'$ is the volume enclosed by the metallic surroundings.
The noise power $P_{\rm b}^{\rm(k)}$ and $P_{\rm zpf}^{\rm(k)}$ can be evaluated using Eq. \ref{Eq2}, \ref{Eq10a},\ref{Eq10b},\ref{Eq10c}, and \ref{Eq13}.

\subsection{Added noise of the amplifiers}
Readout of cavities with high SNR is prototypically achieved using a cascade of amplifiers.
The amplifiers also inevitably introduce added noises.
The total added noise is
\begin{equation}
  P_{\rm a} = P_{\rm a,1} + \frac{P_{\rm a,2}}{G_{1}} + \frac{P_{\rm a,3}}{G_{1}G_{2}} + \dots,
  \label{Eq22}
\end{equation}
where $P_{\rm a,i}$ is the added noise introduced by the i-th amplifier, and $G_{\rm i}$ is the gain of the i-th amplifier.
Apparently the total added noise is dominated by the first-stage amplifier.
When the gains of the amplifiers are sufficiently large, $P_{\rm a} \approx P_{\rm a,1}$.

Linear amplifiers such as Josephson parametric amplifiers (JPA) and high-electron-mobility transistor (HEMT) amplifiers are the most widely used at a few GHz.
The noise power of a cascade of such amplifiers is
\begin{equation}
  P_{\rm a}  = G_{\rm H}B[\hbar\omega(\frac{1}{e^{\hbar\omega/k_{\rm B}T_{\rm phys}}-1}+\frac{1}{2})+k_{\rm B}T_{\rm n}],
  \label{Eq23}
\end{equation}
where $G_{\rm H}$ and $T_{\rm n}$ are the gain and the effective noise temperature of the amplifier chain, and $B$ is the frequency bin of the spectrum.
Particularly, the added noise of JPAs can reach the quantum limit, i.e., $k_{\rm B}T_{\rm n} = \hbar\omega/2$.
For dark matter searching, the SNR is optimized as $B$ equals the linewidth of the expected dark matter signal.

\subsection{Noise spectra and effective sensitive width}
\begin{figure}[http]
  \centering
  \includegraphics[width=1\columnwidth]{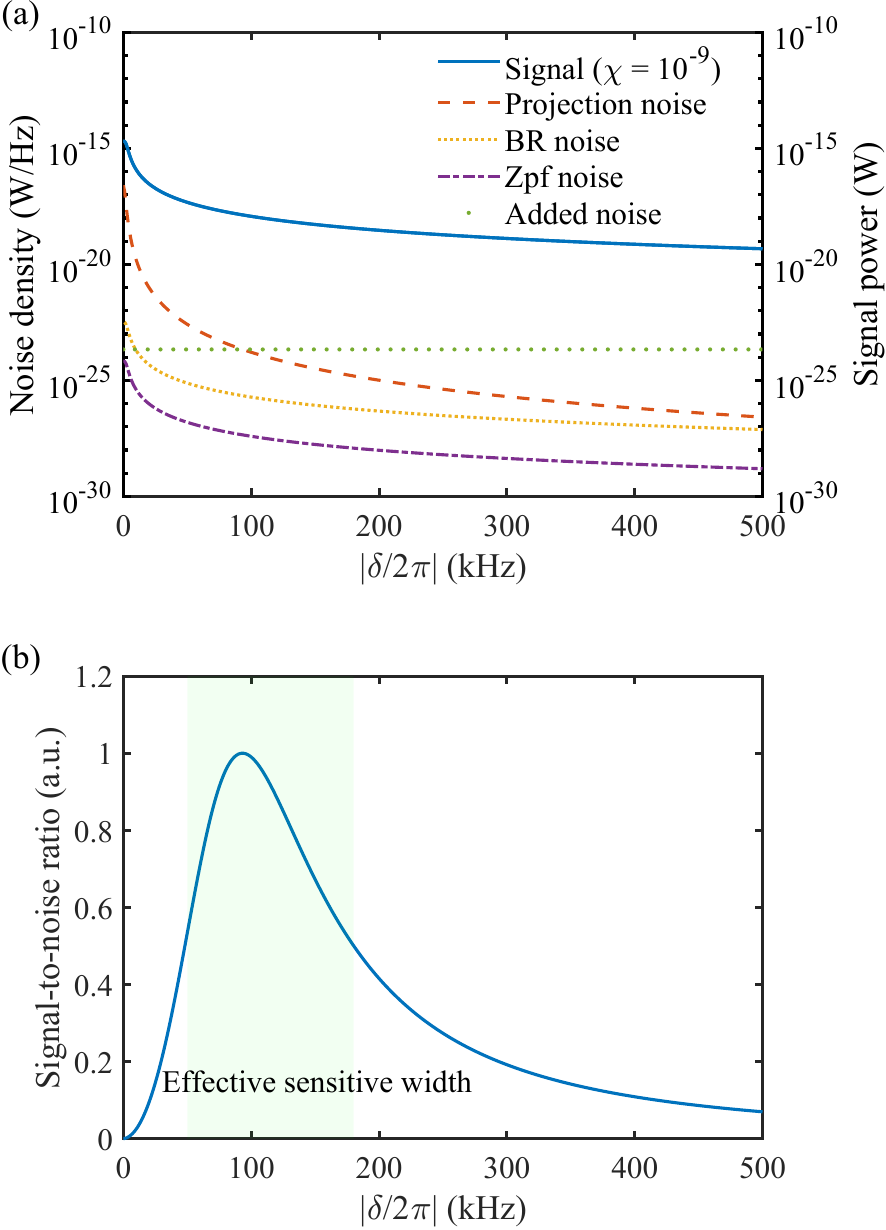}
  \caption{
    (a) Noise spectra contributed by the projection noise (orange dashed line), the black-body radiation noise (yellow densely dotted line),
          the zpf noise (purple dashdotted line),
          and the amplifier added noise (green loosely dotted line).
    (b) The normalized signal-to-noise ratio and the effective sensitive width.}
  \label{FigureS1}
\end{figure}

We take the parameters of Config. 1 in Table 2 of the main text to calculate the noise spectra.
The results are shown in Fig. \ref{FigureS1}(a).
The orange dashed line, the yellow densely dotted line and the purple dashdotted line refer to the spectral density of the projection noise, the zpf noise and the amplifier added noise, respectively.
As the frequency detuning is smaller than 100 kHz, the total noise is dominated by the projection noise.
While as the frequency detuning is larger than 100 kHz, the total noise is dominated by the added noise of the amplifiers.
The solid blue line is the expected signal power with a kinetic mixing of $10^{-9}$.
Figure \ref{FigureS1}(b) plots the normalized signal-to-noise ratio (SNR) as the solid blue line.
The SNR is maximized at $|\delta/2\pi|=100$ kHz.
The shallow green region in Fig. \ref{FigureS1}(b) refers to the effective sensitive width.
In this region the signal-to-noise ratio remains approximately the same order.

\section{Calibration}

\subsection{Gain and noise of the amplifiers}

\begin{figure}[htp]
  \centering
  \includegraphics[width=1\columnwidth]{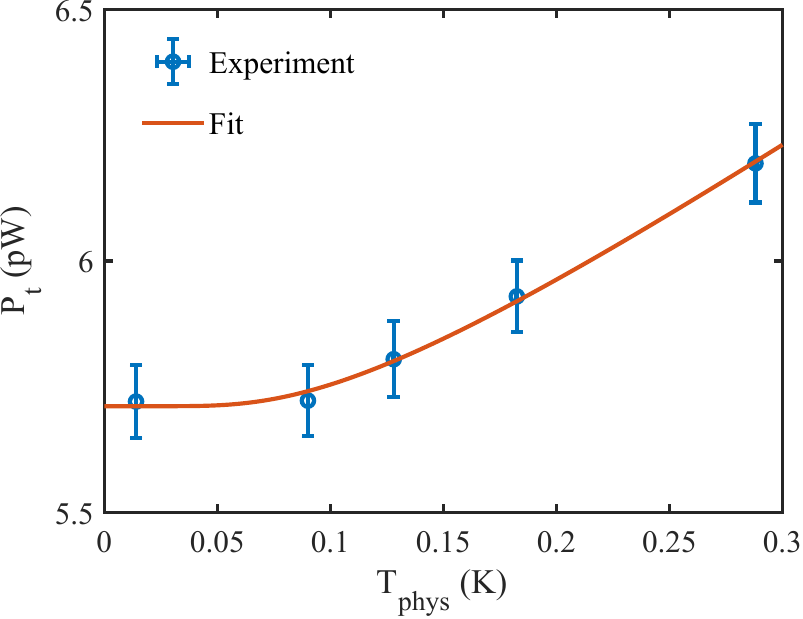}
  \caption{
    Temperature dependence of the added noise power of the amplifiers.
    The y-axis and x-axis refer to the thermal noise power and the temperature in the dilution refrigerator, respectively.
    The experimental and fitting results are shown as the blue dots with error bars and the orange lines, respectively.}
  \label{FigureS3}
\end{figure}

In order to determine the gain and the noise temperature of the amplifier chain, the noise powers were measured at varying physical temperatures $T_{\rm phys}$.
The blue dots in Fig \ref{FigureS3} are the experimental data.
The results are fitted to Eq. \ref{Eq23}, as the orange line in Fig \ref{FigureS3} shows.
The fitting gives $G_{\rm H}=86.7 \pm 0.4$ dB and $T_{\rm n} = 1.7\pm0.1$ K.

\subsection{Package coefficient}

The package coefficient is estimated as
\begin{equation}
  \kappa = \kappa_{1}\kappa_{2},
  \label{Eq25}
\end{equation}
where $\kappa_{1}$ describes the shielding effect of the ground metal around the experimental qubits and the sapphire substrate, and $\kappa_{2}$ is introduced by the sample cell.
The simulation result gives $\kappa_{1} = 0.02$.
The sample cell was a superconducting cylindrical cavity with a radius $R = 30.5$ mm.
Therefore, $\kappa_{2}$ can be calculated as
\begin{equation}
  \kappa_{2} = 1-\frac{J_{0}(0)}{J_{0}(\omega_{\rm q}R/c)},
  \label{Eq26}
\end{equation}
For the experimental qubits $\rm Q_{1}$, $\rm Q_{2}$, and $\rm Q_{3}$, $\kappa_{2}$ is 163, 45, and 15, respectively.

\subsection{The optimal working point}

For the best sensitivity, the driving power of the qubits and the readout frequency should be decided.

\begin{figure}[http]
  \centering
  \includegraphics[width=1\columnwidth]{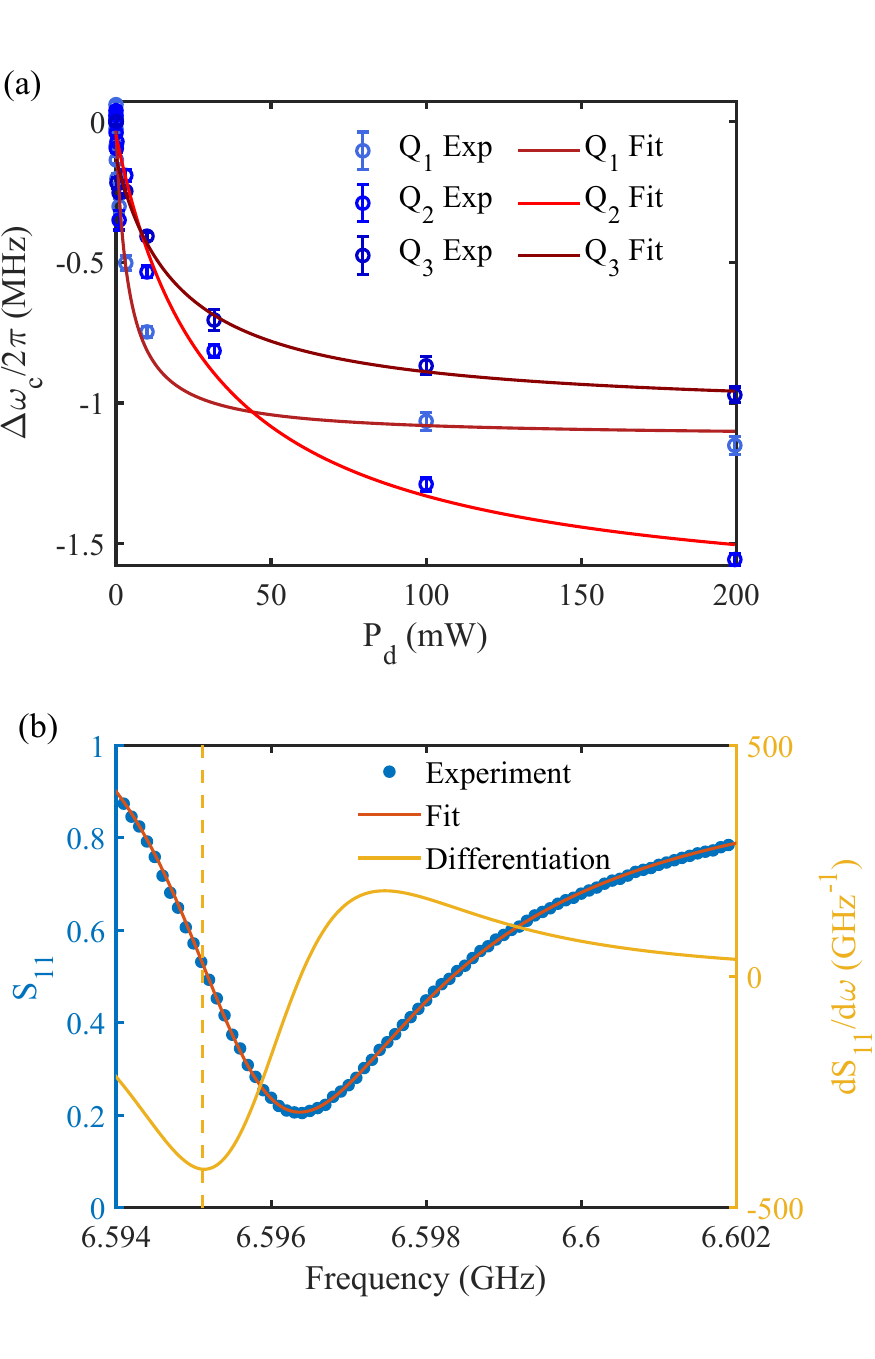}
  \caption{
    (a) Dependence of the resonant frequency of the cavity to the driving power of the qubits.
    (b) The reflection coefficient curve of the cavity at the optimal driving power.}
  \label{FigureS4}
\end{figure}

According to Eq. \ref{Eq10a}, \ref{Eq10b} and \ref{Eq10c}, the best sensitivity is achieved when
  $\Omega_{\rm d} = 0$ and $\gamma = \widehat{\tau}_{\rm z}^{0}/2$.
Eq. \ref{Eq10a} can be modified to
\begin{equation}
  \gamma = \widehat{\tau}_{\rm z}^{0}\frac{1}{1+\mathcal{A}P_{\rm d}},
  \label{Eq27}
\end{equation}
where $\mathcal{A}$ is an arbitrary conversion coefficient and $P_{\rm d}$ is the driving power applied to the qubit.
Consequently, the dispersive shift of the cavity is
\begin{equation}
  \varDelta\omega_{\rm c} = \widehat{\tau}_{\rm z}^{0}g(\frac{1}{1+\mathcal{A}P_{\rm d}} - 1).
  \label{Eq28}
\end{equation}
We measured the bias of the resonant frequency of the cavity with the qubits resonantly driven at varying power.
The experimental results are shown as the royal blue ($\rm Q_{1}$), blue ($\rm Q_{2}$), and medium blue ($\rm Q_{3}$) dots in Fig. \ref{FigureS4}(a).
By fitting the data to Eq. \ref{Eq28}, the conversion coefficient $\mathcal{A}^{\rm(Q1)} = 252\ \rm W^{-1}$, $\mathcal{A}^{\rm(Q2)} = 32\ \rm W^{-1}$, $\mathcal{A}^{\rm(Q3)} = 50\ \rm W^{-1}$, and the coupling strength $G^{\rm(Q1)}/2\pi = 55\pm 1$ MHz, $G^{\rm(Q2)}/2\pi = 68\pm 3$ MHz, $G^{\rm(Q3)}/2\pi = 49\pm 3$ MHz are obtained.
Consequently, the optimal driving powers are $P_{\rm d,op}^{\rm(Q1)} = 4$ mW, $P_{\rm d,op}^{\rm(Q2)} = 31$ mW, and $P_{\rm d,op}^{\rm(Q3)} = 20$ mW,
The fitting results are shown as the firebrick ($\rm Q_{1}$), red ($\rm Q_{2}$), and dark red ($\rm Q_{3}$) lines in Fig. \ref{FigureS4}(a).
As the qubits were continuously driven by a resonant microwave with the optimal power, the $S_{11}$ curve of the cavity was measured, as the blue dots in Fig. \ref{FigureS4}(b).
This curve can be fitted to Eq. \ref{Eq12}.
The orange and yellow lines in Fig. \ref{FigureS4}(b) refer to the fitting result and its first-order differentiation, respectively.
The fitting result gives $Q_{\rm L} = 1755\pm 1$ and $\beta = 0.4229\pm 0.0003$.
The optimal readout frequency is chosen as the frequency where the absolute slope of the $S_{11}$ curve is maximized, as indicated by the yellow dashed line at $\omega_{\rm c}' = 2\pi\times 6595.120$ MHz in Fig. \ref{FigureS4}(b).

\begin{widetext}
\subsection{Flux tuning of the qubits}

In order to tune the frequencies of the qubits, DC currents were injected from $\rm F_{1,2,3}$ Ports. The current generated magnetic flux through the SQUIDs in the qubits, whose Josephson inductance is magnetic-flux-dependent, and thus changed the eigenfrequencies of the qubits.
The current dependence of the eigenfrequencies of the qubits are shown in Fig \ref{FigureS5}.
\begin{figure}[htp]
  \centering
  \includegraphics[width=1\columnwidth]{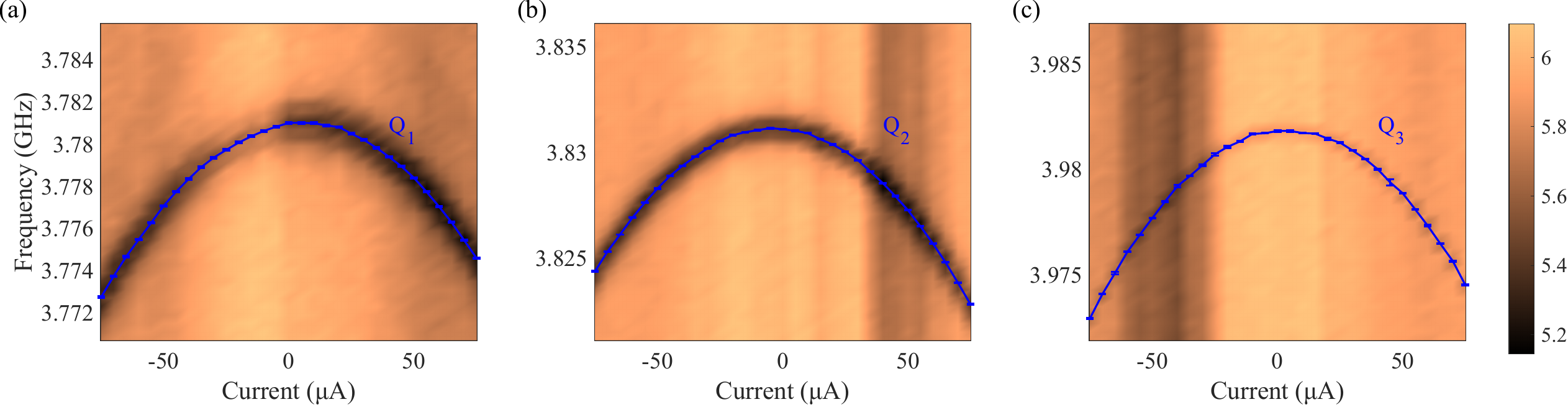}
  \caption{
    Current dependence of the eigenfrequencies of (a) Qubit 1, (b) Qubit 2, and (c) Qubit 3.
    }
  \label{FigureS5}
\end{figure}
\begin{figure}[htp]
  \centering
  \includegraphics[width=1\columnwidth]{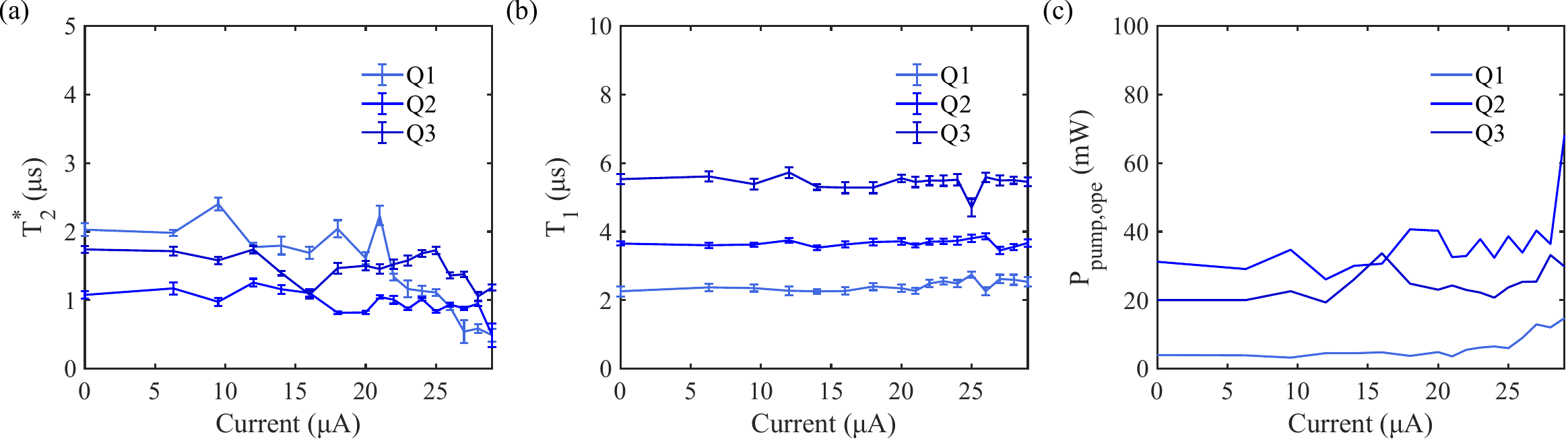}
  \caption{
    Current dependence of (a) the transverse relaxation time, (b) the longitudinal relaxation time, and (c) the optimal driving power.
    }
  \label{FigureS6}
\end{figure}
During the experiment, the current was tuned from 0 µA to 29 µA, so that each qubit swept through a frequency range of around 1 MHz.
The current dependence of the transverse relaxation time, the longitudinal relaxation time and the optimal driving power of the qubits are shown in Fig. \ref{FigureS6}
\end{widetext}
\section{Data analysis}

\begin{figure}[htp]
  \centering
  \includegraphics[width=1\columnwidth]{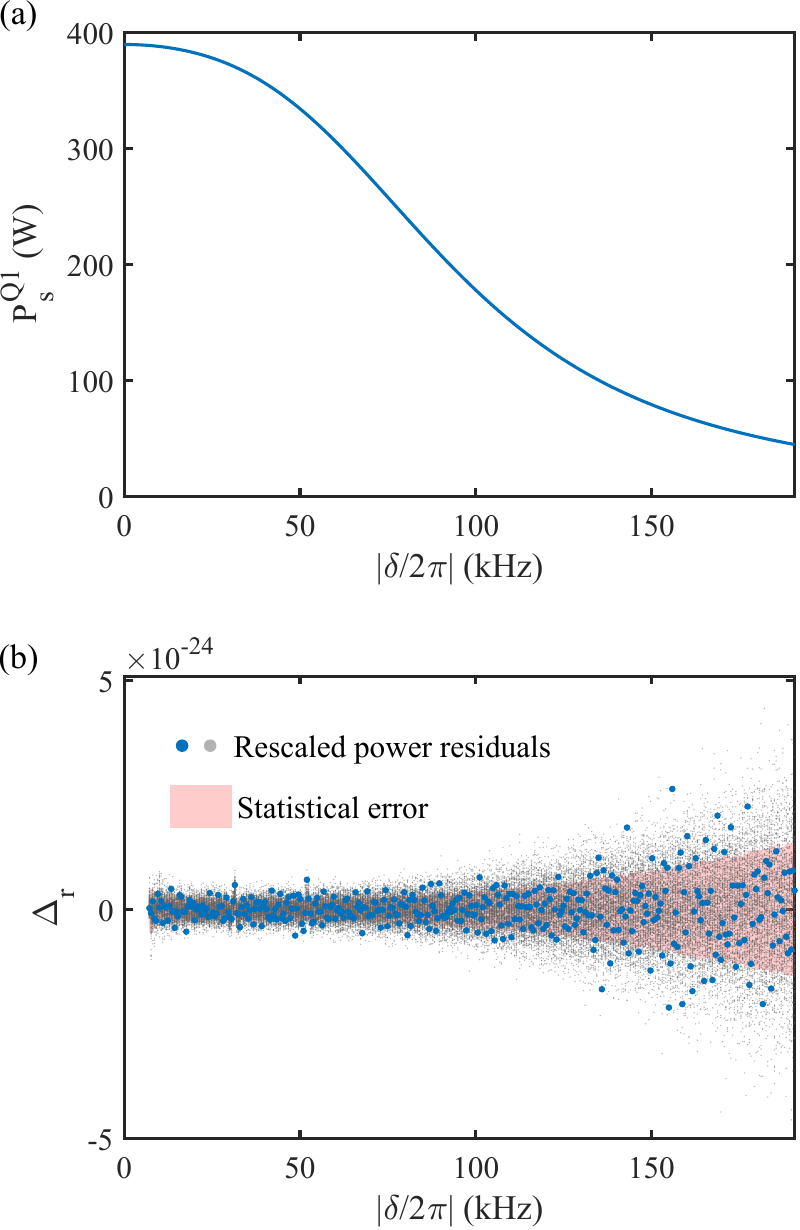}
  \caption{(a) The response function. (b) The rescaled power residuals.}
  \label{FigureS7}
\end{figure}

If a dark photon signal with a frequency close to the eigenfrequency of the first qubit $\rm Q_{1}$ existed, its amplitude would depend on its frequency detuning.
To rescale the potential dark photon signal to a unified amplitude and enable further process, all power residual spectra should be divided by the gain of the amplifier cascade and the response function of $\rm Q_{1}$, i.e., the expected dark photon signal as $\chi = 1$.
Figure \ref{FigureS7}(a) shows the response function, $P_{\rm s}^{\rm Q1}(\chi=1)$.
The rescaled power residuals
\begin{equation}
  \Delta_{\rm r} = \frac{\Delta_{\rm s}\sigma}{G_{\rm H}P_{\rm s}^{\rm Q1}(\chi=1)}
  \label{Eq29}
\end{equation}
is plotted in Fig. \ref{FigureS7}(b).

\begin{figure}[htp]
  \centering
  \includegraphics[width=1\columnwidth]{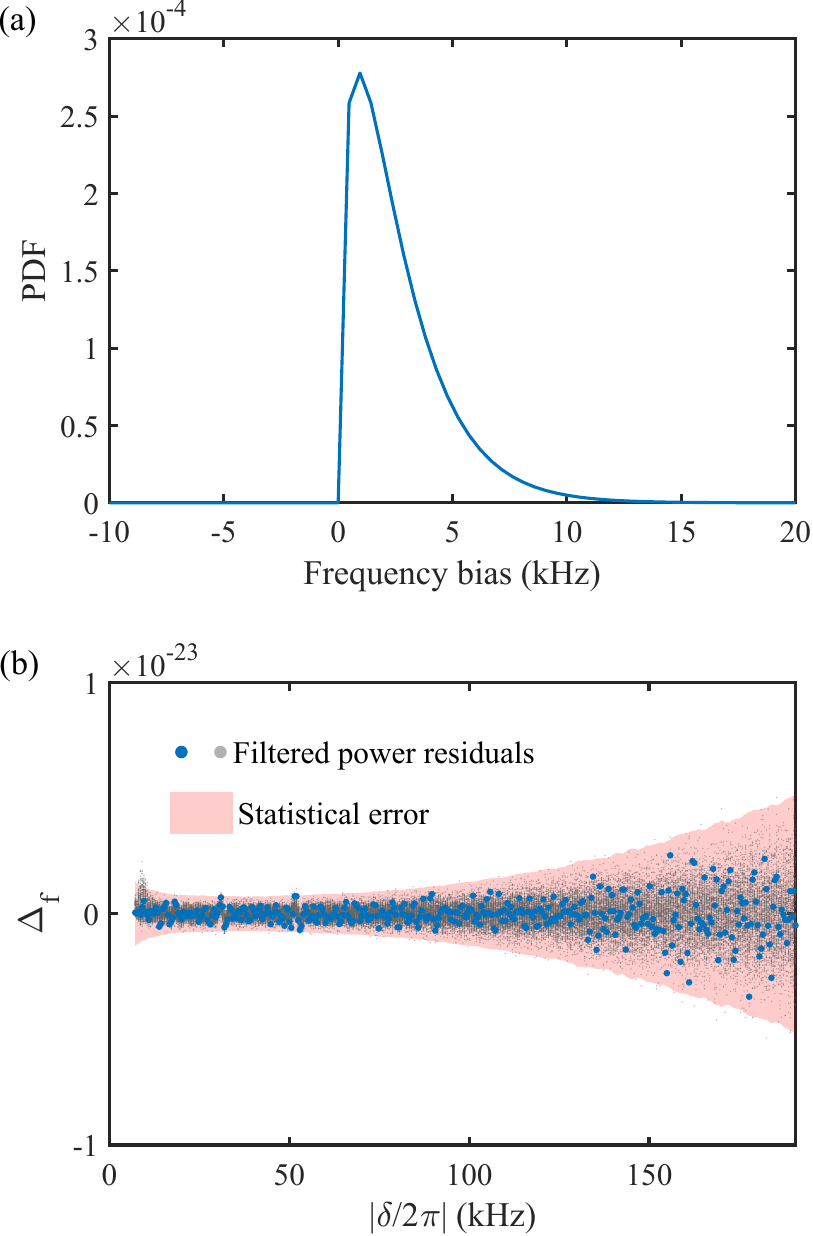}
  \caption{(a) The expected line shape of dark photons. (b) The filtered power residuals.}
  \label{FigureS8}
\end{figure}

The searching efficiency is optimized when the frequency bin width $B$ of the spectrum equals the linewidth of the dark photon.
However, in practice, $B$ is usually smaller than the linewidth of the dark photon.
Consequently, the expected dark photon signal would spread over several frequency bins and the SNR would be degraded.
In order to recover the optimal SNR, a convolution with the line shape of dark photons, which is shown in Fig. \ref{FigureS8}(a), is applied to the rescaled power residuals.
Figure \ref{FigureS8}(b) shows the filtered power residuals $\Delta_{\rm f}$ obtained from the convolution.

\begin{figure}[htp]
  \centering
  \includegraphics[width=1\columnwidth]{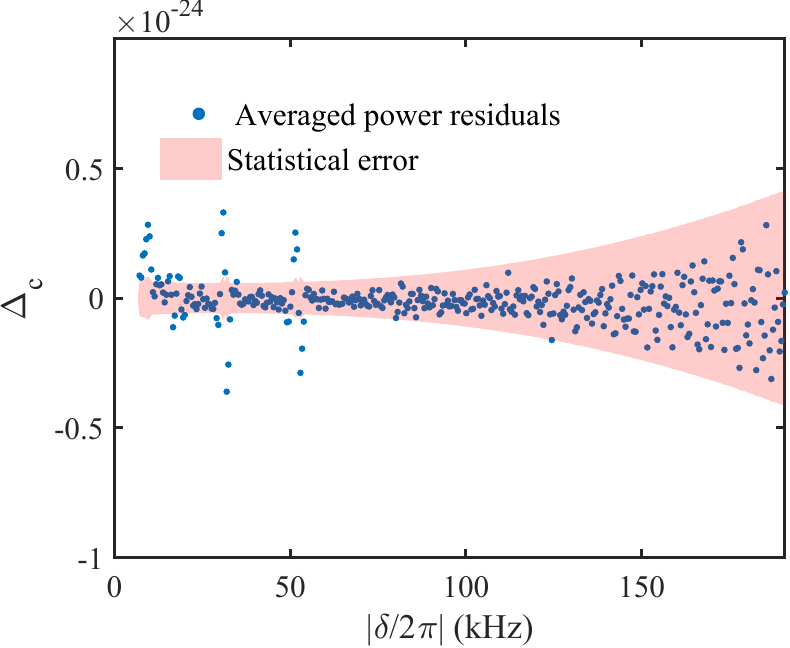}
  \caption{The combined power residuals.}
  \label{FigureS9}
\end{figure}

All the filtered power spectra are combined into a single spectrum to further suppress the noises.
The maximum-likelihood estimation of the combined power residuals $\Delta_{\rm c}$ and the uncertainty $\sigma_{\rm c}$ are calculated as
\begin{equation}
  \Delta_{\rm c} = \frac{\sum_{i = 1}^{150}\frac{\delta_{\rm f;i}}{\sigma_{\rm f;i}^{2}}}{\sum_{i = 1}^{150}\frac{1}{\sigma_{\rm f;i}^{2}}},\ \ \ \
  \sigma_{\rm c} = \sqrt{\frac{1}{\sum_{i = 1}^{150}\frac{1}{\sigma_{\rm f;i}^{2}}}},
  \label{Eq30}
\end{equation}
where $i$ is the index of the subsets.
Systematic errors are introduced by analyzing the uncertainties of the experimental parameters.
The systematic errors are combined with the statistical errors as
  $\sigma'_{\rm c} = \sqrt{\sigma_{\rm c}^{2}+\Delta_{\rm c}^{2}\sum_{x}\sigma_{x}^{2} }$,
where $\sigma_{x}$ stands for the relative uncertainty of parameter $x$.
The combined power residuals and the total errors are shown as the blue dots and the red ribbons in Fig. \ref{FigureS9}, respectively.
It is noteworthy that, although there are several high power residuals, they did not exist through the total experiment, and their line shapes are inconsistent with a dark photon signal.
Therefore, at the frequencies of these power residuals, constraints on the kinetic mixing can also be set.

Bayesian analysis is employed to calculate the constraints on the kinetic mixing $\chi$.
Given a power residual $\Delta_{\rm c}$ and its uncertainty $\sigma'_{\rm c}$, the conditional distribution of $\chi^{2}$ is
\begin{equation}
  \begin{aligned}
    p(\chi^{2}|\Delta_{\rm c}) &= \frac{p(\Delta_{\rm c}|\chi^{2})}{\int_{0}^{+\infty} p(\Delta_{\rm c}|\chi^{2})\,d\chi^{2}},
    \label{Eq31}
  \end{aligned}
\end{equation}
where the distribution of $\Delta_{\rm c}$ is in a Gaussian form,
\begin{equation}
  p(\Delta_{\rm c}|\chi^{2}) = \frac{1}{\sqrt{2\pi}\sigma_{\rm c}}\exp[-\frac{(\Delta_{\rm c}-\chi^{2})^{2}}{2\sigma_{\rm c}^{'2}}].
  \label{Eq32}
\end{equation}
Therefore, the solution of the equation
\begin{equation}
    \int_{0}^{\chi_{90\%}^2} p(\chi^{2}|\Delta_{\rm c}) d\chi^{2} = 90\%
    \label{Eq33}
\end{equation}
gives the constraints on the kinetic mixing with a confidence level of 90\% around $\omega_{\rm q1}$.
Furthermore, by replacing $\rm Q_{1}$ in Eq. \ref{Eq29} by $\rm Q_{2}$ and $\rm Q_{3}$, the constraints on the kinetic mixing around $\omega_{\rm q2}$ and around $\omega_{\rm q3}$ can also be obtained.

\section{Searching for axions and axion-like particles (ALPs)}

Axions and ALPs are another type of dark matter candidate with strong motivation.
It is predicted that in an external magnetic field $B_{0}$, the axion and ALP field can also induce an electric field,
\begin{equation}
  E_{\rm X} = g_{a\gamma\gamma}c^{2}B_{0}\sqrt{2\hbar c\rho_{\rm DM}}/\omega_{\rm X},
  \label{Eq24}
\end{equation}
where $g_{a\gamma\gamma}$ is the axion-photon coupling and $c$ is the speed of light in vacuum \cite{1987Wilczek}.
Therefore, by applying an external magnetic field, the architecture proposed in this article can also be used to search for axions and ALPs.

\begin{figure}[htp]
  \centering
  \includegraphics[width=1\columnwidth]{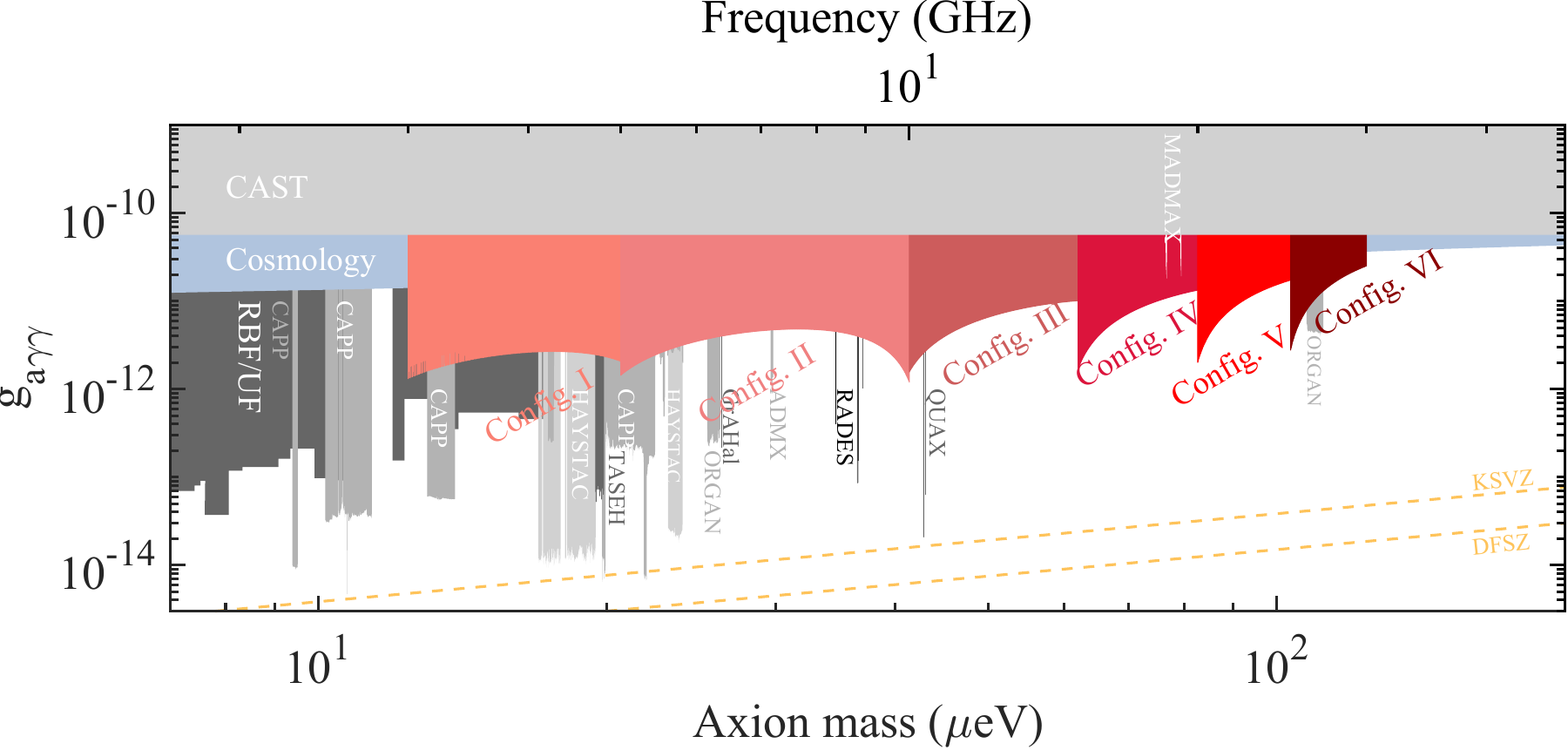}
  \caption{
    The expected exclusion limits on the axion-photon coupling.
    The salmon, light coral, Indian red, crimson, red and dark red regions refer to the expectation of Config. \R{1}, \R{2}, \R{3}, \R{4}, \R{5}, and \R{6} in Table \ref{TableS1}, respectively.
	  For each configuration, the integration time is set to 200 days.
    Data are adapted from \cite{2020Hare}.
    }
  \label{FigureS2}
\end{figure}

Recently, a magnetic-field-compatible transmon qubit has been experimentally realized \cite{2025Gunzler}.
In a magnetic field of 1.2 Tesla, the transverse and longitudinal relaxation time of the qubit achieved 0.6 µs and 6 µs, respectively.
Taking the parameters listed in Table \ref{TableS1}, we can obtain the expected constraints on the axion-photon coupling in the frequency range of 3$\sim$30 GHz, corresponding to the mass range of 12$\sim$66 µeV, as Fig. \ref{FigureS2} shows.

Notably, expected constraints presented in Fig. \ref{FigureS2} are a conservative estimation.
Higher magnetic field can further promote the searching for axions and ALPs.
In 2021, a Josephson junction has been reported to survive a magnetic field of up to 8.5 T \cite{2021Dvir}.
Therefore, qubits made of such Josephson junctions working in strong magnetic field will help to greatly improve the limits on axion-photon coupling.

  \begin{table}[htp]
    \centering
    \renewcommand\arraystretch{1}
    \begin{tabular}{c | c c c c c c}
        \hline
        {Config.}              &{\R{1}}&{\R{2}}&{\R{3}}&{\R{4}}&{\R{5}}&{\R{6}}\\
        \cline{1-7}
        {$\omega_{\rm q}/2\pi$ (GHz)}         &{3$\sim$5}&{5$\sim$10}&{10$\sim$15}&{15$\sim$20}&{20$\sim$25}&{25$\sim$30}\\
		    {$\varpi_{\rm q}/2\pi$ (MHz)}         &{0.5     }&{0.5      }&{0.4       }&{0.4       }&{0.5       }&{0.6}\\
		    {$\varDelta\omega_{\rm q}/2\pi$ (MHz)}&{10      }&{10       }&{8         }&{8         }&{10        }&{12}\\
		    {$Q_{c}$}                    &\multicolumn{6}{c}{$2\times 10^{4}$}\\
		    {$T_{1}$ (µs)}                    &\multicolumn{6}{c}{6}\\
		    {$T_{2}^{*}$ (µs)}                &\multicolumn{6}{c}{0.6}\\
		    {$G/2\pi$ (MHz)}                   &\multicolumn{6}{c}{100}\\
		    {$C$ (fF)}                        &\multicolumn{6}{c}{100}\\
		    {$d$ (µm)}                        &\multicolumn{6}{c}{300}\\
		    {$T_{\rm phys}$ (mK)}             &\multicolumn{6}{c}{10}\\
        \hline
    \end{tabular}
    \caption{Parameters of the detection system.}
    \label{TableS1}
\end{table}

\end{document}